\documentclass[twocolumn,aps,pra]{revtex4-1}

\usepackage{tikz-cd}
	\usepackage{amsmath}
	\usepackage{amsfonts}
	\usepackage{amssymb}
	\usepackage{graphicx}
	\usepackage[colorlinks=true,citecolor=blue,linkcolor=red]{hyperref}
	\usepackage{bbold}					
	\usepackage[makeroom]{cancel}		
	\usepackage{multirow}				
	\usepackage[normalem]{ulem}        
	\usepackage{array}

	\newcolumntype{x}[1]{>{\centering\let\newline\\\arraybackslash\hspace{0pt}}p{#1}}

	\DeclareMathOperator{\Conf}{Conf}  
	
	\DeclareMathAlphabet{\mathbbold}{U}{bbold}{m}{n}
	\def\bs#1{\boldsymbol{#1}}			
	\def\imi{\mathrm{i}}				
	\def\e#1{\mathrm{e}^{#1}}				
	\def\de{\mathrm{d}}
	
	\def\eps{\varepsilon}					
 
	\def\mcH{\mathcal{H}}					
	\def\mcT{\mathcal{T}}					
	\def\mcP{\mathcal{P}}					

	\def\intg{\mathbbold{Z}}					
	\def\ztwo{\mathbbold{Z}_2}					
	\def\zn{\mathbbold{Z}_n}
	\def\triv{\mathbbold{0}}					
	\def\unit{\mathbbold{1}}					
	\def\reals{\mathbbold{R}}					
	\def\cmplx{\mathbbold{C}}					
	
	\def\U#1{\mathrm{U}({#1})}
	\def\GL#1{\mathrm{GL}({#1})}

	\def\bra#1{\left<{#1}\right|}				
	\def\ket#1{\left|{#1}\right>}				
	\def\Herm{\textrm{Herm}}
	\def\Conf{\textrm{Conf}}
	\def\UConf{\textrm{UConf}}
	\def\Disc{\textrm{Disc}}
	\def\U{\textrm{U}}
	\def\GL{\textrm{GL}}
	\def\M{\textrm{M}}
	
	\def\SN{\mathbb{S}_N}
	\def\BN{\mathbb{B}_N}
	\def\PN{\mathbb{P}_N}

	\newcounter{subeqn} %
	\makeatletter
	\@addtoreset{subeqn}{equation}
	\makeatother

\definecolor{TB}{rgb}{1,0.5,0}
\definecolor{XQ}{rgb}{0,0,1}
\definecolor{CW}{rgb}{.5,0,.5}

\begin{document}
\title{Topological Classification of Non-Hermitian Hamiltonians}

\author{Charles C. Wojcik$^1$}
\author{Xiao-Qi Sun$^{2,3}$}
\author{Tom\'{a}\v{s} Bzdu\v{s}ek$^{2,4,5}$}
\author{Shanhui Fan$^1$}\email[Corresponding author: ]{shanhui@stanford.edu}
\
\affiliation{$^{1}$Department of Electrical Engineering, Ginzton Laboratory, Stanford University, Stanford, CA 94305, USA}
\affiliation{$^{2}$Department of Physics, McCullough Building, Stanford University, Stanford, CA 94305, USA}
\affiliation{$^{3}$Stanford Center for Topological Quantum Physics, Stanford University, Stanford, CA 94305, USA}
\affiliation{$^{4}$Condensed Matter Theory Group, Paul Scherrer Institute, CH-5232 Villigen PSI, Switzerland}
\affiliation{$^{5}$Department of Physics, University of Z\"{u}rich, 8057 Z\"{u}rich, Switzerland}

\date{\today}

\begin{abstract}
We revisit the problem of classifying topological band structures in non-Hermitian systems. Recently, a solution has been proposed, which is based on redefining the notion of energy band gap in two different ways, leading to the so-called ``point-gap'' and ``line-gap'' schemes. However, simple Hamiltonians without band degeneracies can be constructed which correspond to neither of the two schemes. Here, we resolve this shortcoming of the existing classifications by developing the most general topological characterization of non-Hermitian bands for systems without a symmetry. Our approach, which is based on homotopy theory, makes no particular assumptions on the band gap, and predicts several amendments to the previous classification frameworks. In particular, we show that the 1D invariant is the noncommutative braid group (rather than $\intg$ winding number), and that depending on the braid group invariants, the 2D invariants can be cyclic groups $\mathbb{Z}_n$ (rather than $\intg$ Chern number). We interpret these novel results in terms of a correspondence with gapless systems, and we illustrate them in terms of analogies with other problems in band topology, namely the fragile topological invariants in Hermitian systems and the topological defects and textures of nematic liquids.
\end{abstract}

\maketitle


\section{Introduction} 


Topological invariants associated with energy bands in the reciprocal momentum ($\bs{k}$-) space have proven useful in predicting novel physical phenomena~\cite{Hasan:2010,Qi:2011}, including robust unidirectional transport, in both electronic and photonic systems.  
Examples of topological invariants include Chern numbers~\cite{Thouless:1982}, which are defined for general systems lacking any particular symmetry, as well as $\mathbb{Z}_2$-invariants~\cite{Kane:2005} and winding numbers~\cite{Asboth:2015}, which are defined as long as some symmetry is preserved. Classification schemes such as the tenfold way~\cite{Horava:2005,Kitaev:2009,Ryu:2010,Kruthoff:2017} provide a unified approach within the mathematical framework of homotopy theory and enable a systematic understanding of the implications of different symmetries for topological invariants.

Non-Hermitian Hamiltonians have widespread applications
in describing open systems. For example, the ubiquitous loss and gain in photonic systems ~\cite{Zhou:2018,Lu:2015,Chen:2016,Noh:2017,Cerjan:2018a,Wang:2017,Zhen:2015,Zhou:2019,Zeuner:2015,Xiao:2017,Hu:2017,Poli:2015,Weimann:2016,Zhan:2017,Choi:2010,Borgnia:2019,Xu:2017,Cerjan:2018, Song:2019b}, the finite quasiparticle lifetimes~\cite{Kozii:2017,Shen:2018,Papaj:2018,Yoshida:2018,McClarty:2019}, and certain statistical-mechanical models~\cite{Hatano:1996}, etc., are naturally described in terms of non-Hermitian Hamiltonians. Recently, there has been a growing interest in uncovering novel topological phases in non-Hermitian systems~\cite{Shen:2018a,Yao:2018,Yao:2018a,Gong:2018,Song:2019,Wang:2019,Budich:2019,Yang:2019,Carlstrom:2018,Okugawa:2019,Moors:2019,Zyuzin:2018,Yoshida:2019,Carlstrom:2019,Bergholtz:2019,Kawabata:2019,Song:2019}. Although these questions have been partially addressed in theory~\cite{Kawabata:2018, Gong:2018,Zhou:2019a}, a unified mathematical description of non-Hermitian band topology is still lacking, even for the most basic setting when no symmetry is assumed.
This is most manifest in the innate dichotomy of the recently suggested classification framework,
which distinguishes two schemes, called the ``line-gap'' resp.~the ``point-gap'' scheme~\cite{Kawabata:2018, Borgnia:2019a}. Within the line-gap scheme, the complex energy spectrum is assumed to miss a line in the complex plane.
This allows one to deform the Hamiltonian into one which is Hermitian with no symmetries (class A), implying integer topological invariant in even spatial dimensions. In contrast, within the point-gap scheme, the complex energy spectrum is assumed to miss a point in the complex plane.
This facilitates a continuous deformation into a Hermitian Hamiltonian with chiral symmetry (class AIII), and implies integer topological invariants in odd spatial dimensions. 

However, there are interesting topologically non-trivial non-Hermitian Hamiltonians that are not uniquely characterized by either a point gap or a line gap. A prototypical example of a such a Hamiltonian is an exceptional ring~\cite{Kawabata:2019}, which arises when a generic non-Hermitian perturbation is applied to a Weyl point~\cite{Wan:2011, Cerjan:2018}. Although our focus in this paper is on gapped systems, the gapless exceptional ring provides a vivid illustration of the difficulty of separately considering point and line gap classifications.
The exceptional ring carries both a 1D and a 2D invariant simultaneously, depending on which type of gap one considers. Curiously, the two invariants have non-trivial influence on each other and therefore cannot always be decoupled. 
Especially, Ref.~\cite{Sun:2019} showed that in  non-Hermitian systems with exceptional lines, the Chern number of 
an exceptional ring ceases to be conserved, but can change sign through a reciprocal braiding process~\cite{Sun:2019,Bouhon:2019,Wu:2018b,Tiwari:2019}.
This observation suggests the need for a more general classification framework, which does not make any assumption on the band gap being globally a point vs.~a line, and which naturally incorporates the interaction between the invariants defined on manifolds with various spatial dimensions.

In this paper, we develop such a unified classification using homotopy theory. 
Within a concise mathematical framework, we revisit the derivation of
1D and 2D topological invariants of non-Hermitian bands. Our derivation
amends the previous theoretical works~\cite{Kawabata:2018, Gong:2018,Zhou:2019a} and explicitly captures the interaction of these invariants. \
Specifically, the theory allows us to classify non-Hermitian Hamiltonians defined on a periodic 2D $\bs{k}$-space, with the surprising result that depending on the braid-group-valued 1D invariants, the $\intg$-valued Chern numbers can be replaced by $\zn$-valued invariants, where the value of $n$ depends on the details of the cycle type of the braids along the Brillouin zone torus.
We note that the interaction of topological invariants in various dimensions is an intrinsically non-Hermitian phenomenon, which is inaccessible in perturbative approaches which start from Hermitian models.
The phenomenon manifests that classical homotopy-theoretic data besides homotopy groups~\cite{Hatcher:2002} play a role in topological band theory beyond the tenfold way. 

The manuscript is organized as follows. The first two sections present the classification result in a detailed, pedagogical manner. First, in Sec.~\ref{sec:derivation-Herm}, we revisit the derivation of the topological invariant for two-band Hermitian systems. The goal is to reformulate this simple story into a mathematical language that is more appropriate for the generalization to non-Hermitian systems, which constitute the contents of Sec.~\ref{sec:derivation-nonHerm}. In this section, we highlight the novelty which arises in the non-Hermitian setting, both in terms of rigorous mathematics and intuitive pictures. Next, in Sec.~\ref{sec:motivation}, we describe the interaction between 1D and 2D invariants as reported for gapless non-Hermitian systems in Ref.~\cite{Sun:2019}, and explain how this interaction relates to the modified topological classification of gapped systems. 
To further extend the intuition for the new classification, we also report on certain relationships with the physics of nematic liquids~\cite{Volovik:1977,Mermin:1979,Alexander:2012} and with the fragile topology of real-symmetric Bloch Hamiltonians~\cite{Wu:2018b,Tiwari:2019,Bouhon:2019,Ahn:2019,Ahn:2019b, Bouhon:2019b}, which provide a useful analogy for understanding the topology of non-Hermitian Bloch Hamiltonians. We also provide a calculation of the new invariants in an explicit model. Finally, in Sec.~\ref{sec:many-band}, we generalize these results to many bands, finding braid group and $\mathbb{Z}_N$ invariants.

\section{Topological classification of two-band Hermitian Hamiltonians}~\label{sec:derivation-Herm}


For electron bands (Hermitian), the notion of topology has been prominent in the last decades. This notion emerges naturally from the ultimate task of condensed matter physics, which is to classify and discover phases of matter and study phase transitions. At the phase-transition critical point, many systems show scale-invariant characteristics, which indicates that there is no finite characteristic length or energy scales at low energy. For a non-interacting electron problem, this point corresponds to a gapless band-structure. Therefore, a phase transition corresponds to a gap closing process and the notion of topology is essential to describe phases: as long as the continuous tuning of the Hamiltonian does not result in a gap closing, the system remains in the same phase. The equivalence class of Hamiltonians tuning without closing a gap therefore becomes a topology problem.

In this section, we review this classification problem using a two-band example in 2D, in a formalism that can be generalized to non-Hermitian cases. We split our presentation into four subsections. In the first subsection, we introduce notation and define the classification problem. The main objects of interest which we introduce are a topological space $X$ (here it is the space of Hermitian Hamiltonians with a spectral gap) and a set $[T^2, X]$ (equivalence classes of such Hamiltonians defined on a Brillouin zone torus $T^2$). In the second subsection, we develop a characterization of the space $X$. The main result here is that $X$ is homeomorphic to the 2-Bloch sphere $S^2$, i.e. $X \sim S^2$. In the third subsection, we compute the set $[T^2, X]$ and find $[T^2, X] = \mathbb{Z}$. This is done in several steps, the first of which is computing the homotopy groups $\pi_n(X)$ (equivalence classes of gapped Hamiltonians defined on an n-sphere $S^n$, representing strong topological invariants of various dimensions). In the final subsection, we define the action of $\pi_1(X)$ on $\pi_2(X)$, which will be of crucial importance in the non-Hermitian setting.

\subsection{Defining the classification problem}
For simplicity, we consider a two-band Hamiltonian describing a 2D lattice model. In momentum space, the Bloch Hamiltonian is simply a family of $2 \times 2$ Hermitian matrices $\mcH(\bs{k})$, where $\bs{k}$ ranges over the wavevectors in the 2D first Brillouin zone. Because $\mcH(\bs{k})$ is periodic in both directions, we can identify opposite edges of the first Brillouin zone and consider the wavevector $k$ as a point in a torus $T^2$. Then $\mcH(\bs{k})$ defines a continuous map $\mcH : T^2 \to \Herm_2(\mathbb{C})$ from the momentum space torus to the set of Hermitian $2\times 2$ matrices. Furthermore, as we motivate from the notion of a topological phase transition, we are interested in the equivalence class of the Hamiltonians upon continuous deformation without gap closing. Therefore, the space we classfiy is the more restricted space $X$, which is the set of \emph{gapped} $2 \times 2$ Hermitian matrices, i.e.~those with distinct eigenvalues. This can be equivalently and concisely formulated by requiring that the \emph{discriminant}, defined as $\Disc (\mcH) = \prod_{i<j} (\lambda_i - \lambda_j)^2$ [where $\{\lambda_i\}_{i=1}^{\dim (\mcH)}$ are the eigenvalues of $\mcH$], is non-vanishing for the Hamiltonian. Therefore we define our target space of gapped Hamiltonians as

\begin{equation}
    X := \{ \mcH \in \Herm_2(\mathbb{C}) : \Disc(\mcH) \neq 0\}.
\end{equation}
Notably, $\Disc(\mcH)$ is a polynomial in the coefficients of $\mcH$, so $X$ is the complement of a hypersurface inside a 4D vector space and can therefore be expected to be topologically interesting. Then $[T^2, X]$ is the set of homotopy classes of Hamiltonians which have distinct energies at every point in momentum space. Topological classification with this choice of $X$ means that topological invariants can change under continuous deformations of $\mcH(\bs{k})$, but specifically only under those which close the spectral gap (which correspond to phase transitions).

\subsection{Characterizing the target space}
We now need to characterize $X$ in a way that makes its topological structure more apparent. We do this by parameterizing $X$ in terms of eigenvectors and eigenvalues, performing an eigen-decomposition. This allows us to describe $X$ in terms of more familiar topological spaces. According to the spectral theorem, the eigenvectors of a Hermitian matrix $\mcH$ constitute columns of a unitary matrix $U \in \U(2)$. 
The eigenvalues $(\lambda_1, \lambda_2)$ are the diagonal entries of a diagonal matrix $\Lambda$. Due to the gap condition, we require $\lambda_1 \neq \lambda_2$, therefore $\Lambda \in \Conf_2(\mathbb{R}) := \{(\lambda_1, \lambda_2) \in \mathbb{R}^2 : \lambda_1 \neq \lambda_2\}$ (the notation $(\lambda_1, \lambda_2)$ refers to a pair of ordered points along the real line). The eigenvalue decomposition of $\mcH$ is $\mcH = U\Lambda U^{-1}$. Thus our parameterization begins with a map $p :  \U(2) \times \Conf_2(\mathbb{R}) \to X$ which is defined by sending a pair $(U, \Lambda) \in  \U(2) \times \Conf_2(\mathbb{R})$ to the Hamiltonian $\mcH = U \Lambda U^{-1}$.

The map $p$ is the starting point for our parameterization, but there are two forms of redundancy which we must account for before we have a one-to-one parameterization of $X$. First, the eigenvectors are only defined up to multiplicaton by a unit complex scalar (the gauge invariance). This defines an action of the group $\U(1) \times \U(1)$ on $\U(2)$, namely multiplying $U$ on the right by a diagonal unitary matrix. Because $\mcH$ is invariant under this group action, we can replace $\U(2)$ with the quotient group ${\U(2)}/{\U(1) \times \U(1)}$. Second, the ordering of the eigenvalues and eigenvectors is not uniquely determined, as long as they are reordered simultaneously. To be precise, if $\sigma$ is the $2 \times 2$ matrix representing the swap permutation, it is easy to verify that $(U, \Lambda) \mapsto (U \sigma, \sigma^{-1} \Lambda \sigma)$ leaves $\mcH$ invariant. This defines an action of the group $\mathbb{Z}_2$ which we must also divide out. By removing these two redundancies, the parameterization of a given Hamiltonian is uniquely defined, so we have the description of $X$ as 
\begin{align}
    X = \Bigg(\frac{\U(2)}{\U(1)\times \U(1)} \times \Conf_2(\mathbb{R})\Bigg) \Bigg / \mathbb{Z}_2
\end{align}
The equals sign here denotes a homeomorphism of topological spaces. Both of the factor spaces are, like $X$, defined by systems of equations, but they are much more familiar in topology. The space ${\U(2)}/{\U(1) \times \U(1)}$ is a classical example of a homogeneous space in Lie theory, and the space $\Conf_2(\mathbb{R})$ arises in connection with the braid group; both play an important role in algebraic topology in the context of classifying spaces~\cite{Hatcher:2002,Whitehead:1978,Arkowitz:2011}.

\begin{figure}[t!]
    \centering
    \includegraphics[width=0.45\textwidth]{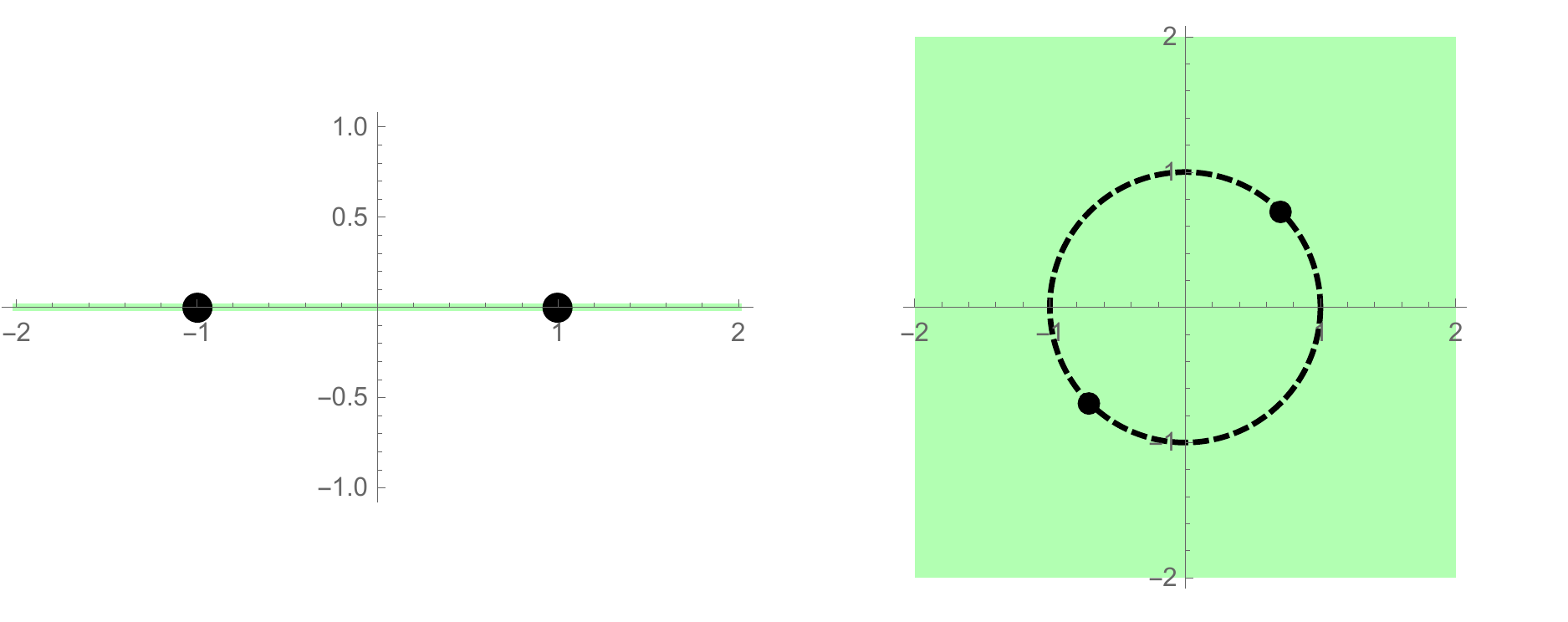}
    \caption{The space of eigenvalues in the Hermitian vs. non-Hermitian case; $\Conf_2(\mathbb{R}) \sim \mathbb{Z}_2$ while $\Conf_2(\mathbb{C}) \sim S^1$.}
    \label{fig:deform}
\end{figure}

The characterization of $X$ we developed is already sufficient for many purposes, but we can further simplify it by performing several improvements. As a first simplification, we recognize the space ${\U(2)}/{\U(1) \times \U(1)}$ as the Bloch sphere $\mathbb{C}P^1 = S^2$. Another simplification we can make (cf.~Fig.~\ref{fig:deform}) is to deform the space $\Conf_2(\mathbb{R}) = \{(\lambda_1, \lambda_2) \in \mathbb{R}^2 : \lambda_1 \neq \lambda_2\}$ into the discrete space $\mathbb{Z}_2 = \{+1, -1\}$ (a form of spectral flattening). Because we can choose the deformation (indicated by the symbol $\sim$) in a way that respects the $\mathbb{Z}_2$ group action, we can retain the parameterization throughout the deformation. We therefore conclude that
\begin{align}
    X &\sim (S^2 \times \mathbb{Z}_2) / \mathbb{Z}_2 \\
    & = S^2.
\end{align}
The intuitive interpretation is that the $S^2$ represents one (e.g.~the lower-energy) eigenvector of the Hamiltonian $\mcH$ on the Bloch sphere.

\subsection{Computing the topological classification}
\begin{figure}[t!]
    \centering
    \includegraphics[width=0.45\textwidth]{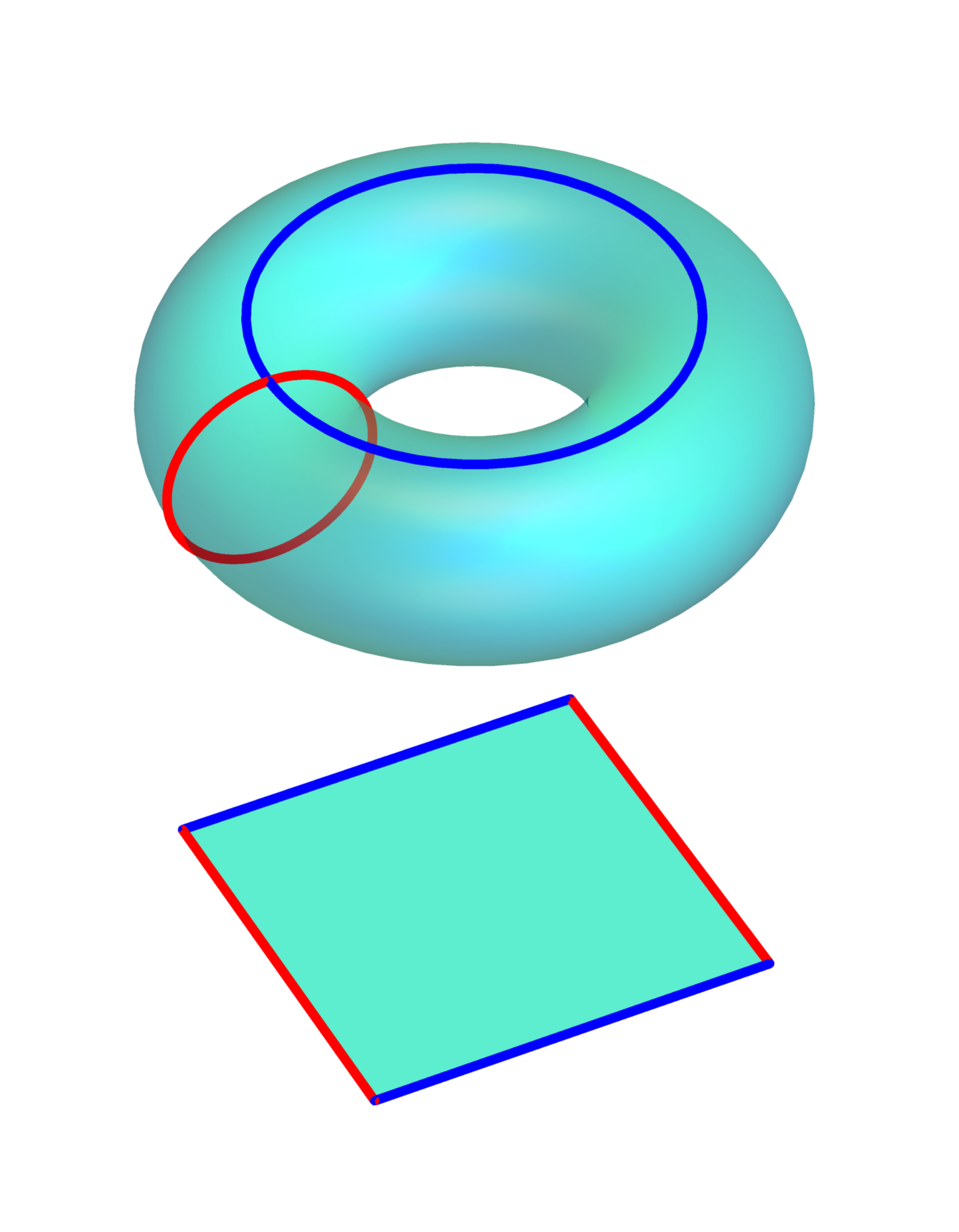}
    \caption{The torus can be constructed out of a rectangle by gluing together the two red lines and the two blue lines. These lines become closed loops on the torus. Taken together, these two loops form the \textit{one-skeleton} of the torus, while the cyan interior of the rectangle forms the \textit{two-cell} of the torus. The one-skeleton $T^2_1 = S^1 \vee S^1$ consists of two circles joined at a common basepoint, i.e. the ``bouquet'' of two circles; we refer to these circles as $a$ and $b$ following the mathematical literature.}
    \label{fig:torus}
\end{figure}

Now that we understand the target space $X$, we are ready to solve the topological classification problem. As shown in Fig.~\ref{fig:torus}, the torus can be thought of as a rectangle with opposite sides identified. The boundary of this rectangle is called the \textit{one-skeleton} of the torus, and it contains information about 1D invariants, while the interior of the rectangle is known as the \textit{two-cell} and contains information about 2D invariants. 

The general strategy is as follows. First, we compute the homotopy groups $\pi_n(X)$, which describe topological invariants of various dimensions; this is a preliminary step which provides data we need to compute $[T^2, X]$. The homotopy groups $\pi_n(X)$ are defined in terms of maps on the $n$-sphere rather than on the torus; we will be interested in $S^1$ and $S^2$, since the torus $T^2$ has non-trivial cycles in 1D and 2D. Next, we use this data to compute $[T^2_1, X]$, where $T^2_1$ is the one-skeleton of the torus. Finally, we study extensions to the two-cell of the torus; this is the key step. Notationally, if $f \in [T^2_1, X]$ is a homotopy class of maps on the one-skeleton, we write $[T^2, X]^f$ to denote the set of homotopy classes of extensions of $f$ to the two-cell, i.e. maps in $[T^2, X]$ which restrict to $f$ on the one-skeleton. As a technicality which we elaborate on in Sec.~\ref{sec:action}, we begin by studying pointed homotopy sets, denoted $[T^2,X]^f_*$ (so that for now, all maps and homotopies preserve basepoints).

Following our outlined strategy, we begin by computing the homotopy groups $\pi_n(X) = [S^n, X]_*$ (considering pointed homotopies). For $X = S^2$, the first few are well-known~\cite{Hatcher:2002}:
\begin{align}
    \pi_1(X) &= 0\\
    \pi_2(X) &= \mathbb{Z} \\
    \pi_3(X) &= \mathbb{Z}
\end{align} 
are the first three. These correspond physically to topological invariants in various dimensions. We understand $\pi_1(X) = 0$ as a statement of the fact that there are no one-dimensional topological invariants in Hermitian systems (without additional symmetry protection). On the other hand, $\pi_2(X) = \mathbb{Z}$ is a statement about 2D topological insulators. The fact that this integer invariant is given by the Chern number is slightly subtle, but can be understood e.g.~in terms of the Chern-Weil~theory~\cite{Milnor:1975}. The three-dimensional ``Hopf'' invariant is unstable, 
meaning it does not survive in the presence of additional bands. Nonetheless, it is still of interest in recent works~\cite{Moore:2008,Deng:2013,Kennedy:2016,Alex:2019}.

The classification of maps on the one-skeleton $[T^2_1, X]_*$ is entirely straightforward, since $T^2_1 = S^1 \vee S^1$ is the wedge product (``bouquet'') of two circles, i.e.~two circles joined at a common basepoint (see Fig.~\ref{fig:torus}). Therefore $[T^2_1, X]_* = \pi_1(X) \times \pi_1(X) = \pi_1(X)^2$. But we found that $\pi_1(X) = 0$, so
\begin{align}
    [T^2_1, X]_* = 0.
\end{align}


The upshot of the result that $[T^2_1, X]_* = 0$ is that we can assume our Hamiltonian $H$ is constant on the one-skeleton of the torus (by continuously deforming it). In other words, we can identify the one-skeleton of the torus to a single point. But then we obtain a sphere, so in this case the extension problem is trivial, and we have
\begin{align}
    [T^2, X]_*^0 = \pi_2(X) = \mathbb{Z}
\end{align}
where $0 : T^2_1 \to X$ denotes the constant map. We have the result
\begin{align}
    [T^2, X]_* = \mathbb{Z}
\end{align}
which constitutes a solution to the classification problem in the Hermitian case.

\subsection{Action of $\pi_1(X)$ on $\pi_2(X)$}~\label{sec:action}
Although the obtained classification is complete, there are some important issues regarding basepoints which we have not yet discussed. More precisely, we have computed the \emph{pointed} homotopy set $[T^2, X]_*$ rather than the \emph{free} homotopy set $[T^2, X]$. However, in the considered physical setting there is no reason to prefer a particular basepoint, thus it would be interesting to study how the homotopy class might change as the basepoint changes; in other words, we are really interested in the free homotopy. In this section, we explain why the two sets could in principle be different, and why the difference can be described in terms of an action of $\pi_1(X)$ on $\pi_2(X)$. Furthermore, although the extension problem was trivial in the Hermitian case, it will not be in the non-Hermitian case, so this is a good time to point out the features which are absent in the Hermitian case which make the non-Hermitian setting richer. The action of $\pi_1(X)$ on $\pi_2(X)$ will also play a central role in understanding the general features of this extension problem.

As a central example, we first describe the non-trivial action of $\pi_1(X)$ on $\pi_2(X)$ for $X = \mathbb{R}P^2 = S^2 / \mathbb{Z}_2$ (the sphere with antipodal points identified), which will provide the intuition for the non-Hermitian case. We represent elements of $\mathbb{R}P^2$ interchangeably as lines or as ellipsoids with a single axis of rotational symmetry, either of which can be thought of as unit vectors with opposite directions identified (due to the symmetry the objects possess). We think of an element of $\pi_2(\mathbb{R}P^2)$ as a texture of ellipsoids on the sphere (see Figs.~\ref{fig:skyrmion}, \ref{fig:extension} for examples of such textures). Starting with a fixed ``Skyrmion'' texture (a generator of $\pi_2(\mathbb{R}P^2) = \mathbb{Z}$), we continuously deform this texture by rotating each ellipsoid in place by $\pi$ radians around the $y$-axis, namely $(x, y, z) \mapsto (-x, y, -z)$. This rotation represents a non-trivial generator of $\pi_1(\mathbb{R}P^2)$; indeed, on $S^2$, this rotation does not create a non-trivial closed loop, but when one identifies antipodal points as in $\mathbb{R}P^2$, it becomes the simplest way to form a non-contractible closed loop at the basepoint. The result of the rotation in the target space $\mathbb{R}P^2$, where $(x, y, z)$ is identified with $(-x, -y, -z)$, is equivalent to a mirror operation across the $xz$-plane in the source space $S^2$, namely $(x, y, z) \mapsto (x, -y, z)$. Thus we clearly see how a generator of $\pi_1(X)$ can turn the initial ``Skyrmion'' texture $1 \in \pi_2(\mathbb{R}P^2) = \mathbb{Z}$ into its mirror image ``anti-Skyrmion'' texture $-1 \in \pi_2(\mathbb{R}P^2)$. Evidently, this could not be done if the target space were $S^2$. 

To more rigorously define the action of $\pi_1(X)$ on $\pi_2(X)$, we use the theory of covering spaces. This theory is developed in Ref.~\cite{Hatcher:2002}. The main theorem we need concerns the homotopy groups of a covering space $Y$ with covering map $p : Y \to X$. It states that the map induces an isomorphism on all $\pi_n$ for $n > 1$, and an injection on $\pi_1$. Intuitively, the covering space is ``unwrapping'' some portion of the $\pi_1$ while leaving the rest of the homotopy unchanged. In fact, there is a one-to-one correspondence between (equivalence classes of) connected covering spaces and (conjugacy classes of) subgroups of $\pi_1(X)$. The universal covering space $\tilde{X}$ is the covering space corresponding to the trivial subgroup. There is an action of $\pi_1(X)$ on any covering space $Y$ by deck transformations; these are defined by lifting a loop $\gamma$ in $X$ to a path $\tilde{\gamma}$ in $Y$ with one endpoint fixed, and seeing where the other endpoint went.

We define the action of $\pi_1(X)$ on $\pi_2(X)$ in terms of the action of $\pi_1(X)$ on the universal cover $\tilde{X}$ by deck transformations. An element $\gamma \in \pi_1(X)$ acts on $\tilde{X}$ by deck transformations, inducing an automorphism $\gamma_*$ of $\pi_2(\tilde{X})$. Using the isomorphisms in $\pi_2$ coming from the covering map $\tilde{X} \to X$, we get an induced automorphism $\gamma_*$ of $\pi_2(X)$. The action of $\pi_1(X)$ on $\pi_2(X)$ is thus defined as a map
\begin{align}
    \rho : \pi_1(X) \to \textrm{Aut}(\pi_2(X)).
\end{align}

Now we can see the role of basepoints in defining $[T^2, X]$. We found $[T^2, X]_* = \pi_2(X) = \mathbb{Z}$ if one allows only pointed homotopies. The only difference if one allows free homotopies is that a free homotopy could incorporate a nontrivial action of $\pi_1(X)$ on $\pi_2(X)$. Thus $[T^2, X]$ is the set of orbits of $[T^2, X]_*$ under the action of $\pi_1(X)$. In the Hermitian case, $\pi_1(X) = 0$, so we have
\begin{align}
    \rho &= 0 \\
    [T^2, X] &= \mathbb{Z}.
\end{align}

We discuss the nontrivial case $\mathbb{R}P^2$ more below. For now, we note that $S^2 \to \mathbb{R}P^2$ is a double cover, and $\pi_1(\mathbb{R}P^2) = \mathbb{Z}_2$ acts on $S^2$ via the antipodal map (the deck transformation in this setting) and thus on $\pi_2(S^2)$ by parity; it follows from the above discussion that
\begin{align}
    \rho(\gamma) = (-1)^\gamma.
\end{align}

\section{Classification of two-band non-Hermitian systems}~\label{sec:derivation-nonHerm}

Now that we have seen the structure of the argument applied to the more familiar Hermitian systems, we can see precisely what changes when one adapts it to non-Hermitian systems. Instead of the space of Hermitian matrices $\Herm_2(\mathbb{C})$, we start by considering the space of all $2\times 2$ matrices $\M_2(\mathbb{C})$. In the non-Hermitian case, several different gap conditions have been considered, leading to differing results~\cite{Kawabata:2018,Borgnia:2019a}. The condition we consider here is natural from a mathematical perspective and results in a classification theory which unifies and extends the existing results. 

We define our target space of gapped non-Hermitian Hamiltonians as 
\begin{align}
    X = \{ \mcH \in \M_2(\mathbb{C}) : \Disc(\mcH) \neq 0\},
\end{align}
i.e.~the space of $2\times 2$ matrices with non-degenerate eigenvalues. To make clear the relationship to the point-gap and line-gap classifications that have previously been studied, note that we are considering \emph{independently} at each wavevector $\bs{k}$ whether or not the complex eigenvalues of $\mcH$ coincide as points in the complex plane. This is a weaker constraint than used by the point-gap scheme, which considers Hamiltonians whose spectrum misses a point (such as  $0$) in the complex plane. It is also very different from the line-gap scheme, which considers Hamiltonians whose spectrum misses a line (such as the imaginary axis) in the complex plane. Under these two schemes, it has been found~\cite{Kawabata:2018} that non-Hermitian Hamiltonians with a point gap have a $\mathbb{Z}$-invariant in dimension one and none in dimension two, while those with a line gap have a $\mathbb{Z}$-invariant in dimension two and none in dimension one.

\subsection{Characterizing the target space}~\label{sec:nonh-target}
With our definition of $X$ in terms of the local gap condition, we need to parameterize $X$ as before in order to compute $[T^2, X]$. The structure of the argument is the same as before; here, we highlight only the relevant differences.

Note that we can again perform an eigen-decomposition, just as we did in the Hermitian case. Indeed, a non-Hermitian matrix with non-degenerate eigenvalues is diagonalizable (e.g.~from the theory of the Jordan normal form~\cite{Lang:2002}). We only have two modifications to consider. First, the eigenvectors are no longer orthogonal, so our matrix of eigenvectors is $G \in \GL_2(\mathbb{C})$ instead of $U \in \U(2)$. Second, the eigenvalues can be complex, so the space $\Conf_2(\mathbb{R})$ is replaced with the space $\Conf_2(\mathbb{C})$, i.e.~the configuration space of ordered pairs of distinct points in the complex plane.

In the Hermitian case, we have argued that the description has a $\U(1) \times \U(1)$ redundancy as well as a $\mathbb{Z}_2$ redundancy. In the non-Hermitian case, the $\U(1) \times \U(1)$ redundancy in the definition of the eigenvectors becomes a $\GL_1(\mathbb{C}) \times \GL_1(\mathbb{C})$ redundancy (recall $\GL_1(\mathbb{C}) = \mathbb{C}^\times$, i.e.~the complex plain without the origin). The $\mathbb{Z}_2$ redundancy remains unchanged. Therefore we have a first description of our target space
\begin{align}
    X = \Bigg(\frac{\GL_2(\mathbb{C})}{\GL_1(\mathbb{C})\times \GL_1(\mathbb{C})} \times \Conf_2(\mathbb{C})\Bigg) \Bigg / \mathbb{Z}_2
\end{align}

As before, while this expression is sufficient for many calculations, we can simplify it to make a clearer picture of the novelties in the non-Hermitian setting. The first factor $\GL_2(\mathbb{C})/\GL_1(\mathbb{C})\times \GL_1(\mathbb{C})$, corresponding to the eigenvectors, is homotopy equivalent to ${\U(2)}/{\U(1) \times \U(1)} = S^2$. This is expected because of the well-known Gram-Schmidt procedure which deforms $\GL_2(\mathbb{C})$ into $\U(2)$, but the actual proof is more complicated; see Ref.~\cite{Crooks:2017}. The second factor, $\Conf_2(\mathbb{C})$, is evidently more interesting than the $\Conf_2(\mathbb{R})$ encountered in the Hermitian case. As we see in Fig.~\ref{fig:deform}, $\Conf_2(\mathbb{C}) \sim S^1$, where a single loop around $S^1$ corresponds to the pair of eigenvalues winding around each other once before returning to their original positions (with the same ordering). Thus we can already see the combination of one- and two-dimensional structure in our characterization
\begin{align}
    X = (S^2 \times S^1) / \mathbb{Z}_2.
\end{align}
This is a key result and will guide our understanding in later sections.

\subsection{Computing the topological classification}~\label{sec:nonh-class}
\begin{figure}[t!]
    \centering
    \includegraphics[width=0.45\textwidth]{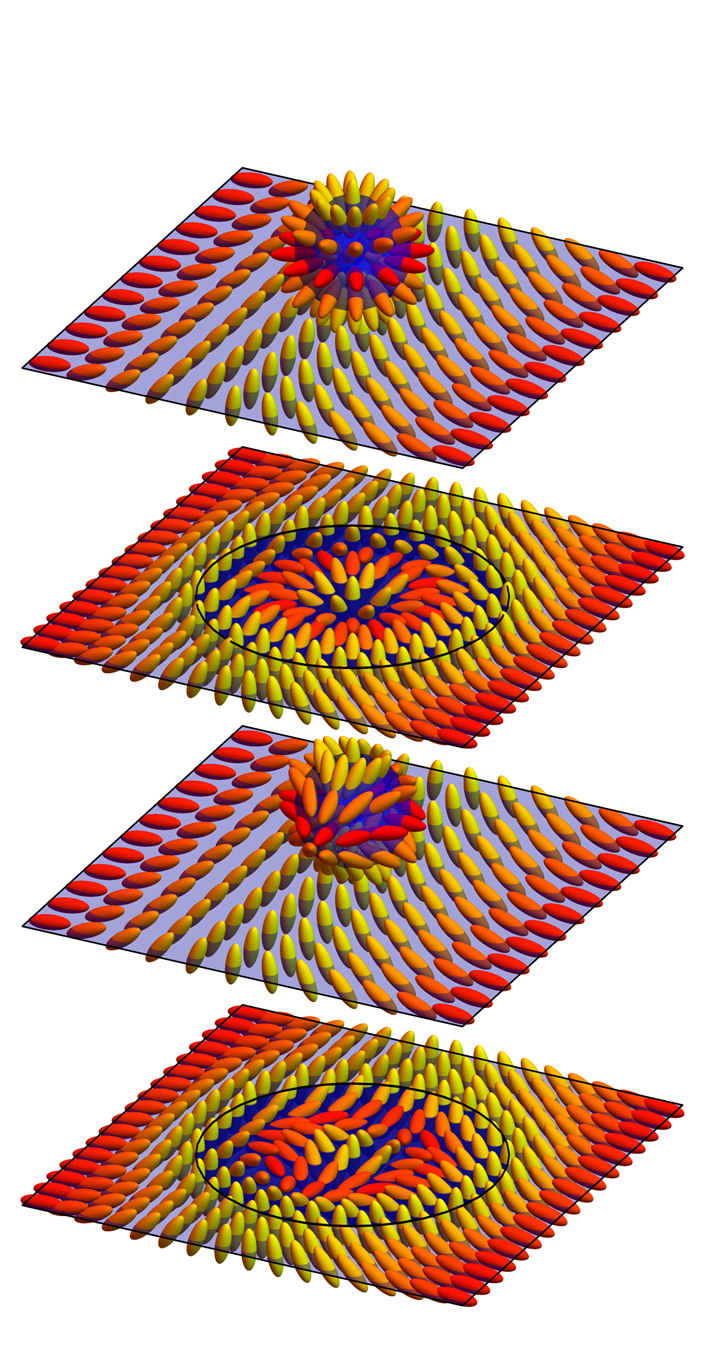}
    \caption{A texture on the sphere is merged with a texture on the torus to form a new texture on the torus. Two configurations are shown, a Skyrmion and and anti-Skyrmion.}
    \label{fig:skyrmion}
\end{figure}
Now that we have a simple characterization of the space $X$, we can begin to study the topological classification. The homotopy groups are easily obtained from the description $X = (S^2 \times S^1) / \mathbb{Z}_2$. Indeed, $X$ has a double cover $Y = S^2 \times S^1$ which has $\pi_1(Y) = \mathbb{Z}$ and $\pi_2(Y) = \mathbb{Z}$. From general properties of double covers (Sec.~\ref{sec:action}), the covering map $Y \to X$ induces an isomorphism in $\pi_2$ and ``unwraps'' $\pi_1$ to some extent. To be precise, $\pi_1(Y) = \mathbb{Z} \subset \pi_1(X)$, and the quotient is $\pi_1(X) / \pi_1(Y) = \mathbb{Z}_2$.
This is consistent with $\pi_1(X) = \mathbb{Z}$, which we verify in Appendix~\ref{sec:fiberbundles} (and $\pi_1(Y)$ sits inside as the even integers). To gain some insight here, we consider the projections $\Pi_2 : X \to S^2/\mathbb{Z}_2 = \mathbb{R}P^2$ and $\Pi_1 :  X\to S^1/\mathbb{Z}_2 = S^1$. The map $\Pi_2$ selects the unordered eigenvectors and the map $\Pi_1$ selects the unordered eigenvalues (after deformations) of $\mcH$.
As we verify in Appendix~\ref{sec:fiberbundles}, $\Pi_1$ induces an isomorphism on $\pi_1$, so $\pi_1(X) = \mathbb{Z}$. On the other hand, $\Pi_2$ induces an isomorphism on $\pi_2$ and reduction mod two on $\pi_1$ ($\pi_1(\mathbb{R}P^2) = \mathbb{Z}_2$).
So we have the homotopy groups $\pi_n(X)$, namely
\begin{align}
    \pi_1(X) &= \mathbb{Z}, \\
    \pi_2(X) &= \mathbb{Z}
\end{align}
and all the higher homotopy groups agree with those of the sphere $S^2$. We see that all the novelty in the non-Hermitian case originates in the eigenvalue winding in 1D. However, as we discuss below, this drastically changes the topological classification in the higher dimensions as well.

\begin{figure}[t!]
    \centering
    \includegraphics[width=0.45\textwidth]{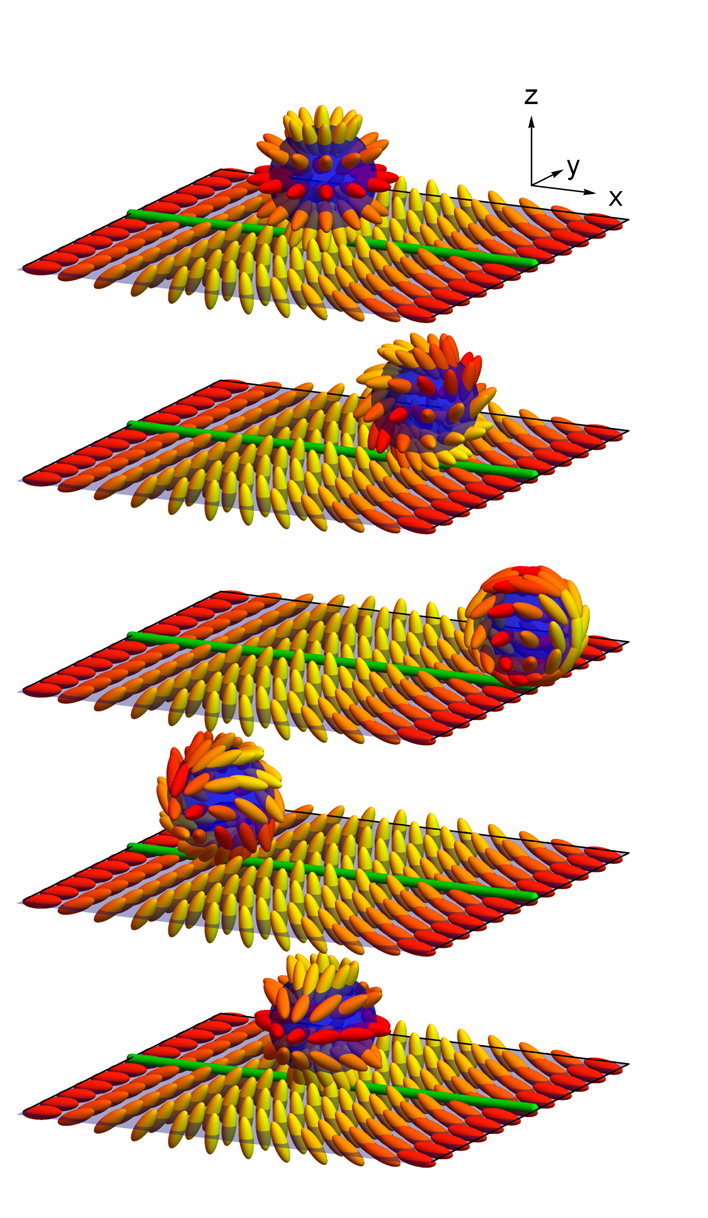}
    \caption{An $\mathbb{R}P^2$ texture on a sphere and a torus, drawn as a field of ellipsoids; the figure illustrates the mechanism by which the action of $\pi_1(X)$ on $\pi_2(X)$ leads to a reduction from a $\mathbb{Z}$ invariant to a $\mathbb{Z}_2$ invariant on the torus. The texture on the torus has nontrivial winding in the $x$ direction, while the texture on the sphere has nontrivial ``Chern number''. The sphere is moved around a complete cycle in the $x$-direction, and meanwhile each ellipsoid undergoes a $\pi$-rotation around the $y$-axis. In the end, the texture on the sphere is equivalent to the texture on the $xz$-mirror of the original sphere. The color indicates the angle of rotation around the $y$-axis.} 
    \label{fig:extension}
\end{figure}

The calculation of the homotopy groups also gives some insight into their nature. The space $\Conf_2(\mathbb{C})/\mathbb{Z}_2$ is also known as the unordered configuration space $\UConf_2(\mathbb{C})$. The one-dimensional invariant is given by the winding of the eigenvalues in $\UConf_2(\mathbb{C})$ (so they are allowed to swap after a complete cycle). The two-dimensional invariant of $X$ comes from the unordered eigenvectors as an element of $\mathbb{R}P^2$. Moreover, since $S^2 \to \mathbb{R}P^2$ induces an isomorphism in $\pi_2$, we see that any map $S^2 \to \mathbb{R}P^2$ can be lifted to a map $S^2 \to S^2$. In other words, for a family of Hamiltonians parameterized by a sphere, one can consistently choose a global ordering of the complex eigenenergies. Then the two-dimensional $\mathbb{Z}$-invariant is just the ordinary Chern number.

As an additional conceptual simplification, it is convenient to think of $X$ as a ``proxy'' space which closely resembles $\mathbb{R}P^2$. Formally, this is because the map $X \to \mathbb{R}P^2$ induces isomorphisms on $\pi_m$ for $m > 1$ and is the reduction modulo 2 ($\mathbb{Z} \to \mathbb{Z}_2$) on $\pi_1$, and therefore for our purposes remembers all important homotopy-theoretic data. This analogy facilitates visualization of the following calculation, and leads to particular physical insights outlined in Sec.~\ref{sec:models} below.


Now we are ready to study $[T^2, X]_*$. We use the same approach as in the Hermitian case. First, we find for the one-skeleton that
\begin{align}
    [T^2_1, X]_* = \mathbb{Z}^2.\label{eqn:stuff}
\end{align}
This is an intuitive result, since we have a pair of integers describing the eigenvalue winding in each direction on the torus. Now fixing $f : T^2_1 \to X$, we need to compute $[T^2, X]_*^f$. We identify $f$ with $(a, b) \in \mathbb{Z}^2$ describing its winding in both directions, and write $f = (a, b)$. Recall that in the Hermitian case, because the map $f$ was trivial on the one-skeleton, we could replace extensions with elements of $\pi_2(X)$. Here, because $f$ is non-trivial, we need a more sophisticated approach. Our approach is inspired by obstruction theory \cite{Arkowitz:2011}, but we present it in elementary terms.

Our approach is to compute $[T^2, X]_*^f$ by defining an action of $\pi_2(X)$ on $[T^2, X]_*^f$, showing that it is transitive (meaning every pair of extensions is related by some group element), and computing the stabilizer of an arbitrary extension $\phi$. Then we will have $[T^2, X]_*^f = \pi_2(X) / \textrm{Stab}_{\pi_2(X)}(\phi)$. For the purpose of these constructions, it is most helpful to visualize elements of $[M, X]$ as $\mathbb{R}P^2$ textures on $M$. The action is defined as in Fig.~\ref{fig:skyrmion} by gluing a small sphere onto the torus. In this figure, we show two configurations, a Skyrmion and an anti-Skyrmion, on $S^2$; later, we will discuss how one can continuously go from one to the other, but for now, we just compare the two textures. Starting with a sphere, one can puncture the sphere and the torus at a point, inflate these points, and glue the resulting boundaries together, resulting in a flattened out version of the texture from the sphere now residing on the torus. In this way, $\pi_2(X)$ acts on $[T^2,X]_*^f$, modifying any given texture on the torus to produce a new texture on the torus with the same behavior on the one-skeleton. The action can be shown to be transitive; for proofs of these claims using cohomology in the context of obstruction theory, see \cite{Ellis:1988, Arkowitz:2011, Ang:2018rls}.

The stabilizer of an arbitrary extension $\phi \in [T^2, X]_*^f$ consists of textures on the sphere which are not homotopic but which become homotopic once glued onto the torus. It turns out that the only way this can occur arises from the action $\rho : \pi_1(X) \to \textrm{Aut}(\pi_2(X))$ \cite{Ang:2018rls}. In Fig.~\ref{fig:extension}, we see the mechanism by which this occurs: a sphere can be moved around a nontrivial cycle on the torus. One way to understand this is that although $\pi_2(X)$ is defined in terms of pointed homotopies, a pointed homotopy on the torus can be realized which results in a free homotopy on glued spheres (making it impossible to consistently choose a lift in $\pi_2(S^2)$). However, not all free homotopies can be realized by moving the sphere around on the torus. The only ones which can be realized are those coming from $[T^2_1, X]_*$ (via the action of $\pi_1(X)$ on $\pi_2(X)$). The mathematical claim is that the stabilizer in $\pi_2(X)$ of $\phi$ is the subgroup generated by elements of the form $s - \rho(f(\gamma)) s$, where $s \in \pi_2(X)$ and $\gamma \in \pi_1(T^2)$. In other words,
\begin{align}
    [T^2, X]_*^{(a, b)} = \pi_2(X) / \langle 1 - \rho(a), 1 - \rho(b) \rangle.
\end{align}
The angle brackets denote the subgroup generated by a collection of elements; the statement is that the elements of $\pi_2(X)$ which become trivial on $T^2$ are precisely those which can be written as linear combinations of the two elements which are obtained by comparing a texture with that obtained by moving it around either the $a$ or $b$ direction.


We know from the preceding discussion that $[T^2, X]_*^f = \pi_2(X) / \langle 1 - \rho(a), 1 - \rho(b) \rangle$, but we haven't yet computed the action of $\pi_1(X)$ on $\pi_2(X)$. Fortunately, this is straightforward from our description $X = (S^2 \times S^1) / \mathbb{Z}_2$. The even subgroup of $\pi_1(X)$ corresponding to the double cover $S^2 \times S^1$ clearly acts trivially on $\pi_2(X)$. The odd subgroup acts via deck transformations on $S^2 \times S^1$, which restrict to the antipodal map on $S^2$. Because this map is orientation-reversing, we see that odd elements of $\pi_1(X)$ act by negation on $\pi_2(X)$. Compare this with Fig.~\ref{fig:extension}, where we illustrate this claim for $\mathbb{R}P^2$. Altogether, we find that $\pi_1(X)$ acts on $\pi_2(X)$ by parity-conditioned negation~\cite{Sun:2019}:
\begin{align}
    \rho(a) = (-1)^a.
\end{align}
We observe that $1 - (-1)^a$ is $0$ for $a$ even and $2$ for $a$ odd. There are four cases for the parity of $a$ and $b$ to consider; in each case, the stabilizer subgroup in $\pi_2(X) = \mathbb{Z}$ is either $0$ or $2\mathbb{Z}$. Therefore
\begin{equation}
    [T^2, X]_*^{(a, b)} = \begin{cases}
    \mathbb{Z} \textrm{ if $a$, $b$ are both even} \\
    \mathbb{Z}_2 \textrm{ otherwise}.
    \end{cases}
    \label{invariant}
\end{equation}



This concludes the calculation of the topological classification in the non-Hermitian setting. To summarize, we have
\begin{align}
    [T^2, X]_* = \bigcup_{(a, b) \in \mathbb{Z} \times \mathbb{Z}} \begin{cases}
    \mathbb{Z} \textrm{ if $a$, $b$ are both even} \\
    \mathbb{Z}_2 \textrm{ otherwise}.
    \end{cases}\label{eqn:T2-X-result}
\end{align}
This should be compared with
\begin{align}
    [T^2, \mathbb{R}P^2]_* = \bigcup_{(a, b) \in \mathbb{Z}_2 \times \mathbb{Z}_2} \begin{cases}
    \mathbb{Z} \textrm{ if $a$, $b$ are both zero} \\
    \mathbb{Z}_2 \textrm{ otherwise}
    \end{cases}
\end{align}
obtained for the ``proxy'' simpler space visualized in Figs.~\ref{fig:skyrmion}, \ref{fig:extension}. Finally, we have
\begin{align}
    [T^2, X] = \bigcup_{(a, b) \in \mathbb{Z} \times \mathbb{Z}} \begin{cases}
    \mathbb{N} \textrm{ if $a$, $b$ are both even} \\
    \mathbb{Z}_2 \textrm{ otherwise}
    \end{cases}
\end{align}
which we can compare with
\begin{align}
    [T^2, \mathbb{R}P^2] = \bigcup_{(a, b) \in \mathbb{Z}_2 \times \mathbb{Z}_2} \begin{cases}
    \mathbb{N} \textrm{ if $a$, $b$ are both zero} \\
    \mathbb{Z}_2 \textrm{ otherwise}.
    \end{cases}
\end{align}

The invariants in Eq.~(\ref{eqn:T2-X-result}) are evidently related to Chern numbers. However, the $\ztwo$ case is somewhat subtle, because Chern number is not defined when a global ordering of the bands is absent. One could attempt to integrate over the double cover of the torus to remedy this, but the result of such integration is always zero, because the contributions from the two sheets cancel each other. Finally, one could simply just integrate over the single torus, ignoring the discontinuity at the boundary. However, such procedure does not produce a quantized result, and hence such integration does not represent a topological invariant. We discuss the proper way to compute the $\ztwo$ invariant later in Sec.~\ref{sec:models}, where we also apply the method to study a simple toy Hamiltonian.


\section{Physical interpretation of classification results}~\label{sec:motivation}

In this section, we develop a physical framework for understanding the classification result obtained in Sec.~\ref{sec:derivation-nonHerm}. First, in Sec.~\ref{sec:Chern-Weyl} we describe a well-known relationship between gapped systems and gapless systems, namely how a Chern number can be described in terms of Weyl points. We then show in Sec.~\ref{sec:pi1-pi2} how these notions are challenged in a non-Hermitian model with two energy bands due to the non-trivial interaction between 1D and 2D topological invariants, and how we can extend the correspondence to this setting using Weyl points and exceptional rings.

To strengthen our intuition about the reduced topological classification on a torus, we discuss in Sec.~\ref{sec:nematic} an analogy with the classification of topological defects and textures in nematic liquids~\cite{Volovik:1977,Mermin:1979,Alexander:2012}, where a closely related phenomenon has been known for a long time~\cite{Volovik:1977}. Similar phenomena have been studied more recently in the context of the fragile topology of real-symmetric Hamiltonians~\cite{Ahn:2019,Tiwari:2019}, which we briefly review in Sec.~\ref{sec:fragile-real}. 
\subsection{Correspondence betwen Chern number and Weyl points in Hermitian systems}~\label{sec:Chern-Weyl}

We revisit in this section certain elementary aspects of band topology in Hermitian systems, before shifting our focus in Sec.~\ref{sec:pi1-pi2} to non-Hermitian systems. 
The prototypical example of a topological invariant in Hermitian band theory is the (first) Chern number, which is an integer number assigned to any 2D closed manifold inside the momentum space~\cite{Hasan:2010}. It is defined as the integral of Berry curvature over the manifold, divided by $2\pi$. Importantly, there is an exact mathematical correspondence which allows us to interpret the Chern number on a given manifold in terms of the Weyl points enclosed \emph{inside} that manifold. To be more precise, recall that Weyl points are point-like degeneracies of a pair of bands inside the 3D $\bs{k}$-space~\cite{Wan:2011}. 
Depending on their chiral charge $\chi=\pm 1$, each Weyl point (WP) acts as either a source or a sink of a $2\pi$-quantum of Berry curvature~\cite{Xiao:2010}. Since Berry curvature has vanishing divergence away from band degeneracies, it follows from Stokes' theorem that the integral of the Berry curvature on the boundary $M = \partial \mathcal{D}$ of any region (domain) $\mathcal{D}$ is quantized to integer multiples of $2\pi$, and the Chern number $c$ is exactly equal to the total charge of the Weyl points in $\mathcal{D}$.
\begin{figure*}[t!]
    \centering
    \includegraphics[width=0.99\textwidth]{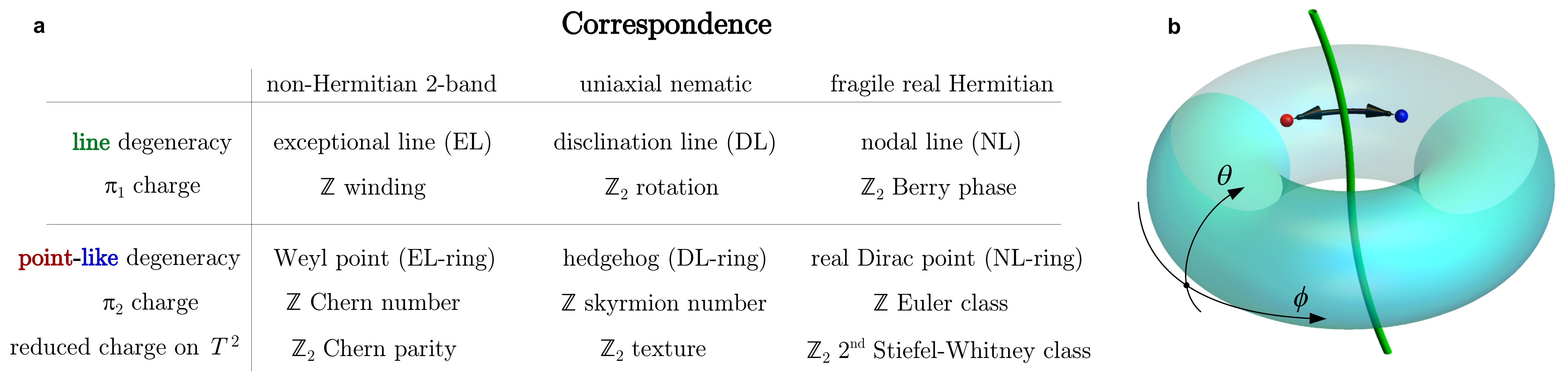}
    \caption{Comparison of the three physical settings discussed in Sec.~\ref{sec:pi1-pi2},~\ref{sec:nematic}, resp.~\ref{sec:fragile-real}. Although the terminology differs, the description of their singularities (topological defects) is very similar. Especially, all three systems generically exhibit line degeneracies in 3D (green line in the illustration), which are protected by non-trivial first homotopy group ($\pi_1$) of the underlying order-parameter space. These line degeneracies can be folded into closed rings, which can be shrunk to point-like degeneracies (red resp.~blue dot in the illustration), either by fine-tuning of the Hamiltonian or due to lowering the energy cost of the defect. If the point degeneracy is robust, it carries an integer charge that follows from non-trivial second homotopy group ($\pi_2$) of the underlying order-parameter space. In each instance, the 1D and 2D invariant interact non-trivially. This means that the $\pi_2$-charge describing a point nodes flips sign (represented by the black arrow) if it is carried along a closed path with non-trivial $\pi_1$ charge (i.e. a path that encloses a line singularity). This interaction leads to  $\intg\longrightarrow \ztwo$  lowering of the topological invariant on the torus, if there is a non-trivial $\pi_1$ charge in some direction. In the illustration, the $\phi$-direction carries a non-trivial $\pi_1$ charge, because it encloses the vertical line defect (green).}
    \label{fig:correspondence}
\end{figure*}

Reformulating the statements mathematically, it follows from the definition of the Chern number
\begin{subequations}\label{eqn:Chern-intro}
\begin{equation}
c^\alpha = \frac{1}{2\pi} \oint_{M} \bs{F}^\alpha(\bs{k})\cdot \de^2\bs{k}\;\;\in\intg\label{eqn:Chern-def}
\end{equation}
that
\begin{equation}
c^\alpha  = \sum_{\textrm{WP}^\alpha \in \mathcal{D}} \chi_\textrm{WP}^\alpha,\label{eqn:Chern-correspondence}
\end{equation}
In Eqs.~(\ref{eqn:Chern-intro}), $\bs{F}^\alpha(\bs{k}) = \imi \bra{\bs{\nabla} u^\alpha_{\bs{k}}}\times\ket{\bs{\nabla} u^\alpha_{\bs{k}}}$ is the Berry curvature on energy band $\alpha$, $\ket{u^\alpha_{\bs{k}}}$ is the corresponding cell-periodic part of the Bloch wave function, and $\bs{\nabla}$ is the gradient operator in $\bs{k}$-space. 
Note that we have fixed one band (labelled $\alpha)$, and we consider only the Weyl points formed by this band. Furthermore, as WPs are degeneracies of pairs of bands, it can be shown~\cite{Xiao:2010} that 
\begin{equation}
\chi_\textrm{WP}^\alpha = -\chi_\textrm{WP}^\beta\quad\textrm{for WP formed by bands $\alpha$ \& $\beta$},\label{eqn:Chern-flip}
\end{equation}
\end{subequations}
meaning that each WP acts as a sink on one of the two bands, and as a source on the other band. Two Weyl points which are both formed by bands $\alpha$ and $\beta$ can annihilate only if their chirality is opposite.

\subsection{Interaction of 1D and 2D invariants in non-Hermitian systems}~\label{sec:pi1-pi2}
The correspondence between topological insulator and band nodes becomes more subtle in a non-Hermitian setting. Although the (first) Chern number has been previously considered in non-Hermitian systems~\cite{Shen:2018a,Yao:2018}, as it is meaningful for some 2D closed manifolds, non-Hermitian Bloch Hamiltonians have an additional 1D invariant~\cite{Shen:2018a} which interacts non-trivially with the Chern number~\cite{Sun:2019} (Sec.~\ref{sec:derivation-nonHerm}). As discussed above, the complication stems from the complex-valued band energies, which allow for the permutation of two energy bands along a closed trajectory without forming a band degeneracy on the way. The presence of such a trajectory inside the 2D Brillouin zone makes it impossible to globally assign each band a unique band index, and therefore Eq.~(\ref{eqn:Chern-def}) cannot be readily applied to compute the first Chern number. 

To get an insight into the nature of the non-Hermitian counterpart of the Chern number, we find it useful to consider again the correspondence with Weyl points. In non-Hermitian systems, it has been found that Weyl points generically turn into one-dimensional ring-like degeneracies known as exceptional rings~\cite{Wan:2011, Cerjan:2018}. These exceptional rings have been found to have nontrivial one-dimensional invariant associated with the winding on a circle threaded by the ring (associated with the point-gap classification scheme) as well as a nontrivial Chern number on a sphere large enough to enclose the entire ring (associated with the line-gap scheme). Inside a torus, a small exceptional ring may be considered effectively as a Weyl point, with the understanding that it will generically have some small but nonzero radius. However, one can consider large exceptional rings which thread through the torus, in either direction (inside or outside).


For simplicity, we consider a two-band model on a donut, see Fig.~\ref{fig:correspondence}\textbf{b}. We assume that the two bands are non-trivially permuted along the $\phi$-direction of the donut, and that originally there are no band degeneracies inside the donut. The non-trivial band permutation is realized by an exceptional nodal line threaded through the donut. Let us now consider the following process: through a local band inversion, we produce a pair of Weyl points or opposite chirality. By appropriately adjusting the Hamiltonian parameters, we transport one of the Weyl points along the $\phi$-direction, while keeping the other Weyl point fixed. Along the path, the transported Weyl point flips upside-down. According to Eq.~(\ref{eqn:Chern-flip}), the flip implies that the Weyl point has effectively reversed the chirality. As a consequence, the two Weyl points now (locally) carry the \emph{same} chirality, and are not able to annihilate anymore. 

Conversely, any pair of Weyl points can be annihilated in a model that exhibits a non-trivial band twist along some direction of a 3D region $\mathcal{D}$. If the two Weyl points locally have the same chirality, one can still annihilate them by transporting one of the Weyl points along the non-trivial path. We thus observe that the right-hand side of Eq.~(\ref{eqn:Chern-correspondence}) is no longer an integer invariant in a non-Hermitian system if there is a non-trivial band flip in some direction of the region $\mathcal{D}$. On the other hand, the \emph{parity} (even vs.~odd) of the total number of Weyl points inside $\mathcal{D}$ remains invariant, as long as no Weyl points are allowed to move across the boundary $\partial \mathcal{D}$, i.e.~as long as $\partial \mathcal{D}$ does not exhibit a gap closing. This change of parity is the manifestation of our $\mathbb{Z}_2$ invariant in terms of the correspondence between gapped and gapless systems.

Thus we understand the classification result Eq.~\ref{eqn:T2-X-result} in terms of Weyl points and exceptional lines / rings inside the torus. While the system is gapped on the torus, it has band degeneracies inside the torus. Exceptional lines / rings which link with the torus are responsible for the 1D part of the classification, i.e. the winding numbers on the one-skeleton; Weyl points are responsible for the 2D part, i.e. the extension to the two-cell. The construction from Sec.~\ref{sec:derivation-nonHerm} of a group action of $\pi_2(X)$ on $[T^2, X]_*^f$ can be understood as the insertion of a Weyl point into the interior of the solid torus. The reduction mod 2 of the $\mathbb{Z}$-invariant under conditions of nontrivial winding is understood in terms of parity-flip of Weyl points (see also Fig.~\ref{fig:extension}).

\subsection{Insights from the physics of nematic liquids}~\label{sec:nematic}

Nematic liquids~\cite{Alexander:2012} are the archetypal 
example of an ordered phase considered in the context of topological defects and textures~\cite{Mermin:1979}. This phase of matter is built up from approximately rod-like molecules, which are randomly positioned (resembling a liquid) but with a frozen orientation (resembling a crystalline solid). The order-parameter of a liquid crystal is the so-called \emph{director}, which is an unioriented axis that describes the local orientation of the molecules. The order-parameter space of such ``uniaxial'' nematics is therefore
\begin{equation}
X = S^2 / \ztwo = \reals P^2 \label{eqn:nematic-RP2}
\end{equation}
where $S^2$ represents a unit vector $\bs{n}$ aligned with the orientation of the molecules, and the quotient identifies $\bs{n}\sim-\bs{n}$ to produce the ``headless'' director. This is exactly the ``proxy'' space considered in detail in Sec.~\ref{sec:derivation-nonHerm}.

The order-parameter field of a nematic liquid in 3D may exhibit topological defects, which can be explained using homotopy groups. On the one hand, the first homotopy group $\pi_1(\reals P^2) = \ztwo$ describes a non-trivial twist of the order-parameter along a closed path ($S^1$). More precisely, this is a $\pi$-rotation of the director, and the corresponding defect is described as a \emph{disclination line}. On the other hand, the second homotopy group $\pi_2(\reals P^2)$ describes a non-trivial texture of the director on a sphere ($S^2$), which is colloquially called \emph{hedgehog}. 

Naively, by recalling the correspondence between Chern number and Weyl points from Sec.~\ref{sec:Chern-Weyl}, one might think 
that nematic textures on a closed surface $\partial \mathcal{D}$ would be characterized by an integer topological invariant that is in correspondence with the number of hedgehogs in $\mathcal{D}$. However, this conclusion is wrong. It has been recognized by Volovik and Mineev~\cite{Volovik:1977} that moving a hedgehog around a disclination line flips its integer topological charge. As a consequence, any pair of hedgehog defects can pairwise annihilate if brought together along a non-trivial trajectory. This is very similar to the way Weyl point chirality in non-Hermitian system is reversed when it is moved around an exceptional line. Therefore, one can draw the following analogy between non-Hermitian systems and nematic liquids:
\begin{subequations}
\begin{eqnarray}
\textrm{exceptional lines}\;&\longleftrightarrow&\;\textrm{disclination lines} \\
\textrm{Weyl points}\;&\longleftrightarrow&\;\textrm{hedgehogs}.
\end{eqnarray}
\end{subequations}
For nematic liquids, it has been found~\cite{Ang:2018rls} that the interaction between the 1D and the 2D invariants reduces the topological classification of textures on torus from $\intg$ to $\ztwo$ whenever there is a non-trivial 1D invariant of the director along some direction of the torus. The $\ztwo$ invariant is in one-to-one correspondence with the parity of the number of hedgehogs inside the torus~cf.~Fig.~\ref{fig:correspondence}\textbf{b}.

\subsection{Fragile topology of real Hermitian models}~\label{sec:fragile-real}

Very recently, the observation that topological invariant on a 1D subspace can reduce the topological classification on a 2D manifold has also been made in the context of topological band theory. More specifically, this phenomenon was reported~\cite{Ahn:2019,Tiwari:2019} for fragile topological invariants~\cite{Po:2018} of models with real-symmetric Hamiltonians. Such condition arises either in the presence of $C_2\mcT$ symmetry (composition of $\pi$-rotation with time reversal) or $\mcP\mcT$ symetry (composition of spatial inversion with time reversal)~\cite{Sjoeqvist_2004,Wu:2018b,Bouhon:2019b}. 

Let us first summarize the so-called ``stable'' topology of such real-symmetric Hermitian models, which correspond to nodal class $\textrm{AI}$ of Ref.~\cite{Bzdusek:2017}. The generic band degeneracy of such Hamiltonians in 3D is a nodal line, protected by a $\ztwo$-valued (quantized) Berry phase on closed paths ($S^1$). Furthermore, nodal lines can be folded to produce closed nodal-line rings, which were reported to carry a $\ztwo$-valued monopole charge~\cite{Fang:2016} on the enclosing sphere ($S^2$). This pair of $\ztwo$ invariants mathematically correspond to so-called first and second Stiefel-Whitney class~\cite{Ahn:2019b,Milnor:1975}. By fine-tuning the Hamiltonian parameters, nodal-line rings with a monopole charge can be shrunk to a point-like degeneracy known as ``real Dirac point''~\cite{Zhao:2017}, resembling 
the way we considered shrinking exceptional nodal-line rings to Weyl points in Sec.~\ref{sec:pi1-pi2}.

In systems with a small number of bands, the groups describing the band nodes of real symmetric Hamiltonians may be enriched. This phenomenon is called \emph{fragile topology}, and its presence for real-symmetric Hamiltonians has been linked~\cite{Song:2019c,Po:2018,Po:2019} to the physics of twisted bilayer graphene near the magic angle~\cite{Bistritzer:2011,Cao:2018a,Cao:2018b}. 
Especially, when such Hamiltonian exhibits two occupied and an arbitrary (but larger than two) number of unoccupied bands, the monopole charge becomes an integer~\cite{Zhao:2017,Bzdusek:2017} called an Euler class~\cite{Ahn:2019,Bouhon:2019,Ahn:2019b}.
It has been reported~\cite{Tiwari:2019} that the Euler class of a nodal-line ring flips sign when it is carried along a closed path with non-trivial Berry phase. As a consequence, the topological classification of real-symmetric Hamiltonians 2D Hamiltonians with two occupied bands reduces from $\intg$ to $\ztwo$ whenever there is a non-trivial Berry phase along some direction of the torus, cf.~Fig.~\ref{fig:correspondence}\textbf{b}. The $\ztwo$ invariant that remains from the integer Euler class is again the second Stiefel-Whitney class, which we mentioned above in the context of the stable topology. One thus finds the following analogy between the non-Hermitian two-band Hamiltonians and the fragile topology of real-symmetric Hamiltonians:
\begin{subequations}
\begin{eqnarray}
\textrm{exceptional lines} \;&\longleftrightarrow&\; \textrm{nodal lines}\\
\textrm{Weyl points} \;&\longleftrightarrow&\; \textrm{real Dirac points}. \\
\textrm{(EL-rings)}\;\;&\phantom{\;\longleftrightarrow\;}&\;\;\textrm{(NL-rings)}\nonumber 
\end{eqnarray}
\end{subequations}
The comparison between the various systems considered in Sec.~\ref{sec:pi1-pi2},~\ref{sec:nematic}, and~\ref{sec:fragile-real} is summarized by the table in Fig.~\ref{fig:correspondence}\textbf{a}.

\subsection{Wilson-loop spectra interpretation of the $\mathbb{Z}_2$ invariant}~\label{sec:models}

Apart from understanding the topological invariants from the view of extending the Brillouin zone to a donut as shown in Fig.~\ref{fig:correspondence}, it is also natural to find an intrinsic 2D algorithm to determine the $\mathbb{Z}_2$ invariant derived in Sec.~\ref{sec:nonh-class} in Eq.~(\ref{invariant}).  Here we find such a geometric interpretation and a corresponding algorithm for the $\mathbb{Z}_2$ invariant. Notice that most of the Hermitian 2D topological invariants appear in some way in the Wilson loop eigenvalue flow~\cite{Soluyanov:2011,Gresch:2017}, so this is a natural place to look for features of the $\mathbb{Z}_2$ invariant. To define the Wilson loop eigenvalue flow, we slice the torus into loops and study the change of Wilson loop eigenvalues along the slicing direction. 

Physically, in the Hermitian case, the Berry phase information in the Wilson loop determines whether the system can be deformed to an atomic insulator. In our non-Hermitian case, we expect such a method still captures the topological invariant. To generalize this to non-Hermitian cases, an essential point to define the Wilson loop is to require a consistent ordering of bands along the loop direction (i.e. no winding). This condition is required for gauge-invariance of the Wilson loop eigenvalues and can be achieved by choosing a proper direction to slice the Brillouin zone torus.

\begin{figure}[!t]
    \centering
    \includegraphics[width=0.48\textwidth]{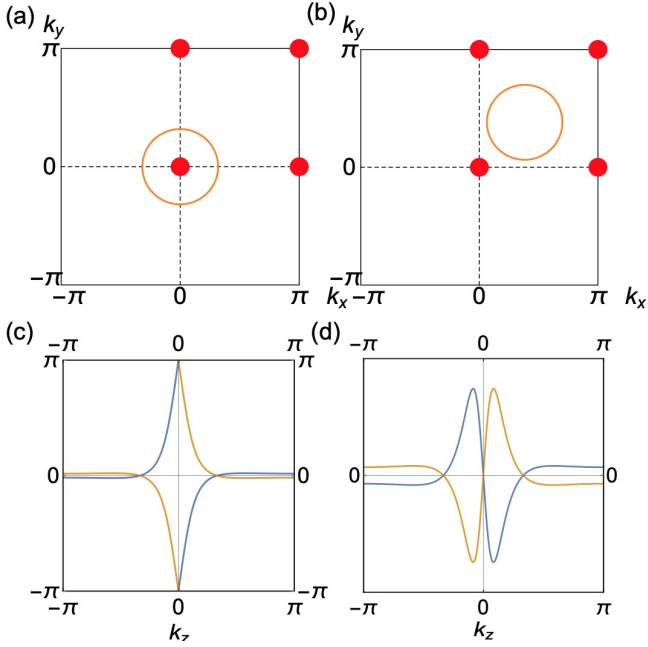}
    \caption{
The flow of Wilson loop eigenvalues of the two bands choosing different cylinders. The parameters used for calculation are $m=2$, $k_r=1$. There are four Weyl points (red) located at $k_z=0$ plane as shown in panels (a) and (b). We choose two different cylinders whose projections are shown as orange loops in panels (a) and (b) for the Wilson loop caculation in panels (c) and (d). Panel (c) shows the flow of Wilson loop eigenvalues when the center of the cylinder lies at $(k_x,k_y)=(0,0)$. The cylinder encloses one Weyl point and the invariant is non-trivial. We see one crossing (odd) at $\pi$ and two crossings (even) at $0$. Panel (d) shows the flow of Wilson loop eigenvalues when the center of the cylinder lies at $(k_x,k_y)=(1.2,1.2)$. The cylinder encloses no Weyl points and the invariant is trivial. We see zero crossing (even) at $\pi$ and three crossings (odd) at $0$. }
    \label{wloop}
\end{figure}

We shall first discuss the $\mathbb{Z}_2$ invariant in the simple case when the winding along one direction $k_x$ of the torus is even and the other direction $k_y$ is odd. We can slice the torus along the $k_y$ direction. For each loop, the energy order is well-defined and we can compute Wilson-loop eigenvalues (Berry phases) for both bands. We use biorthogonal left and right eigenvectors to compute the Berry phases, as is standard for non-Hermitian systems~\cite{Cerjan:2018}.
The winding along the $k_y$ direction requires the two eigenvalues to have odd crossing and switch in between as we vary $k_y$ from $-\pi$ to $\pi$. Furthermore, the two eigenvalues $e^{i\phi_1}$ and $e^{i\phi_2}$ satisfy $\phi_1+\phi_2=0$ mod $2\pi$. Therefore, these crossings can only happen at $\phi_1=\phi_2=0$ or $\pi$. Since the total number of crossings is an odd number, there are two topologically distinct classes of Wilson loop eigenvalue flow: odd number of crossings at $\phi_1=\phi_2=0$ or odd number of crossings at $\phi_1=\phi_2=\pi$. We propose that the two classes correspond to the $\mathbb{Z}_2$ invariant; see Fig.~\ref{wloop} to compare the two cases. We remark that even in the case when the winding is nontrivial in both directions, one can find a slicing direction (e.g. parallel with the diagonal of the Brillouin zone) such that the bands exhibit an even winding in the loop direction~\cite{Ahn:2018}. 

Here we shall test the consistency of the two ways of identifying the $\mathbb{Z}_2$ invariant. We first construct a 3D lattice model~\cite{Sun:2019} with Weyl points and non-trivial winding along one momentum direction:
\begin{eqnarray}
\mcH(\bs{k};m)  
&=& \e{\imi \frac{k_z}{2}}\left[\cos\left(\tfrac{k_z}{2}-\tfrac{\pi}{3}\right)\sin k_x \sigma_x + \right. \nonumber \\ 
&\phantom{=}&+\left.\cos\left(\tfrac{k_z}{2}+\tfrac{\pi}{3}\right)\sin k_y \sigma_y\right. \label{eqn:alt-lattice-model} \\
&\phantom{=}&+ \left. \left(\sin k_z \cos\tfrac{k_z}{2} - 2m \sin \tfrac{k_z}{2} \right)\sigma_z \right]. \nonumber
\end{eqnarray}
The construction is inspired by the correspondence described in Sec.~\ref{sec:pi1-pi2}. For $m>1$, there are four Weyl points at $(k_x,k_y,k_z)=(0,0,0),(\pi,0,0),(0,\pi,0)$ and $(\pi,\pi,0)$. Now we can take a cylinder (more precisely a torus) centered at $k_x=k_y=0$ with certain radius $k_r$, small compared to the separation of the Weyl points. For this choice of cylinder, the total parity of the enclosed Weyl points is $1$. On the other hand, if we shift the center of the cylinder away from $k_x=k_y=0$ to move the Weyl point outside of the cylinder, the total parity of the Weyl points is $0$. We now slice the cylinder along $k_z$ direction and calculate the Wilson loop eigenvalue flow upon changing $k_x$. The results for the corresponding two cases are shown in Fig.~\ref{wloop}(a-b).

\section{General classification of $N$-band models}~\label{sec:many-band}
We now have by now developed a solid understanding of topological invariants in non-Hermitian systems with two bands. Here, we generalize the result to $N$ bands. The structure is even richer, with $\mathbb{Z}$- and $\mathbb{Z}_2$-valued invariants replaced with braid group ($\BN$) and cyclic groups ($\mathbb{Z}_k$) valued invariants. Nevertheless, the overal logic behind the derivation of these results is the same as before.

To start, we use the same gap condition,
\begin{align}
    X = \{\mcH \in \M_N(\mathbb{C}) : \Disc(\mcH) \neq 0\}.
\end{align}
This means that we only consider ``fully gapped'' Hamiltonians in which all the complex eigenvalues are distinct.
The deformations and redundancies from before generalize directly to this scenario, giving
\begin{align}
    X = \Bigg(\frac{\U(N)}{\U(1) \times \ldots \times \U(1)} \times \Conf_N(\mathbb{C})\Bigg) \Bigg / \SN
\end{align}
where the unitary quotient space $U(N)/U(1) \times \ldots \times U(1)$ is known as the complex flag manifold, and $\SN$ is the symmetric group, acting via $N \times N$ permutation matrices. 
In this case, we will not simplify the presentation of $X$ further. Instead, we use well-known results~\cite{Hatcher:2002,Whitehead:1978,Arkowitz:2011} concerning the homotopy groups of the factors, namely
\begin{align}
    \pi_1(\Conf_N(\mathbb{C})) &= \PN \\
    \pi_m(\Conf_N(\mathbb{C})) &= 0, m > 1 \\
    \pi_1(\UConf_N(\mathbb{C})) &= \BN \\
    \pi_m(\UConf_N(\mathbb{C})) &= 0, m > 1 \\
    \pi_1\Bigg(\frac{\U(N)}{\U(1) \times \ldots \times \U(1)}\Bigg) &= 0 \\
    \pi_2\Bigg(\frac{\U(N)}{\U(1) \times \ldots \times \U(1)}\Bigg) &= \mathbb{Z}^{N-1}.
\end{align}
Here, $\PN$ is the pure (ordered) braid group, $\BN$ is the full braid group, and $\Conf_N(\cmplx)$ ($\UConf_N(\cmplx)$) is the ordered (unordered) configuration space of $N$ points in the complex plane.
The first equations are understood in terms of $N$ ordered points in the plane braiding around each other. The last equation describes a Chern number associated to each band, subject to the constraint that the sum over the Chern numbers of all the bands must be zero.

We can compute the homotopy groups of $X$ as before. Since $X$ has a covering space with deck transformation group $\SN$ whose homotopy groups we understand, we obtain immediately $\pi_2(X) = \mathbb{Z}^{N-1}$. We use the projection $X \to \Conf_N(\mathbb{C}) / \SN = \UConf_N(\mathbb{C})$ with simply connected fiber to obtain $\pi_1(X) = \BN$. The interpretation of these invariants is a straightforward extension of the two-band case.

Now we can study $[T^2, X]_*$. On the one-skeleton, we have $[T^2_1, X]_* = \BN^2$. Now $\pi_1(X) = \BN$ acts on $\pi_2(X) = \mathbb{Z}^{N-1}$ by permutations in the standard representation; this is evident from the $\SN$-covering space and the interpretation of $\pi_2(\U(N)/\U(1)\times \ldots \times \U(1))$ as $N$ Chern numbers whose sum is zero.

Let $f : T^2_1 \to X$ be given by a pair of braids $(b_1, b_2)$, and let $(\sigma_1, \sigma_2)$ be the corresponding pair of permutations. We will study extensions of $f$ to the two-cell of the torus by computing the stabilizer for the action of $\pi_2(X)$ on the set of extensions. The relations on $\pi_2(X) = \mathbb{Z}^{N-1}$ are generated by the columns of $\unit - \sigma_1$ and $\unit - \sigma_2$ as matrices in the standard matrix representation. As an example, we study the case $\sigma_1 = (1, \ldots, N)$ is a single $N$-cycle, and $\sigma_2 = 1$ trivial. The notation $\sigma_1 = (1, \ldots, N)$ means that band 1 goes to band 2, band 2 goes to band 3, etc., and band $N$ goes back to band 1. We choose a basis $\{e_i - e_{i+1}\}$ for $\mathbb{Z}^{N-1}$. With respect to this basis, we have
\begin{align}
    1 - \sigma_1 = \begin{pmatrix}
    1 & 0 & 0 &\ldots  & 0 & 0 & 1 \\
    -1& 1 & 0 &\ldots & 0 & 0 & 1\\
    0 &-1 & 1 &\ldots & 0 & 0 & 1\\
    \vdots& \vdots &\vdots & \ddots & \vdots & \vdots & \vdots \\
    0 & 0 & 0 & \ldots & 1 & 0 & 1 \\
    0 & 0 & 0 & \ldots & -1 & 1 & 1\\
    0 & 0 & 0 & \ldots & 0 & -1 & 2
    \end{pmatrix}
\end{align}
Multiplying on the left by the determinant 1 matrix
\begin{align}
    \begin{pmatrix}
    1 & 0 & 0 &\ldots  & 0 & 0 & 0 \\
    1& 1 & 0 &\ldots & 0 & 0 & 0\\
    1 &1 & 1 &\ldots & 0 & 0 & 0\\
    \vdots& \vdots &\vdots & \ddots & \vdots & \vdots & \vdots \\
    1 & 1 & 1 & \ldots & 1 & 0 & 0 \\
    1 & 1 & 1 & \ldots & 1 & 1 & 0\\
    1 & 1 & 1 & \ldots & 1 & 1 & 1
    \end{pmatrix}.
\end{align}
we obtain
\begin{align}
    \begin{pmatrix}
    1 & 0 & 0 &\ldots  & 0 & 0 & 1 \\
    0& 1 & 0 &\ldots & 0 & 0 & 2\\
    0 &0 & 1 &\ldots & 0 & 0 & 3\\
    \vdots& \vdots &\vdots & \ddots & \vdots & \vdots & \vdots \\
    0 & 0 & 0 & \ldots & 1 & 0 & N-3 \\
    0 & 0 & 0 & \ldots & 0 & 1 & N-2 \\
    0 & 0 & 0 & \ldots & 0 & 0 & N
    \end{pmatrix}
\end{align}
whose column space is an index $N$ sublattice of $\mathbb{Z}^{N-1}$. It follows that instead of $N-1$ Chern numbers, the invariant of these $N$ bands reduces to a single $\mathbb{Z}_N$ invariant. The intuition is similar to the 2-band case. The row vector $(1, \ldots, 1)$ provides the map $\mathbb{Z}^{N-1} \to \mathbb{Z}_N$, so we can interpret the $\mathbb{Z}_N$ invariant as the total number of Weyl points $\sum_i e_i-e_{i+1}$ mod $N$ (see Sec.~\ref{sec:models}).  The result for general permutations is more complicated, but can be worked out on a case-by-case basis by computing the Smith normal form~\cite{Carlsson:2009}.

\section{Conclusion}
We have presented a novel topological classification scheme for gapped non-Hermitian systems, which generalizes existing schemes and finds new types of topological invariants. In particular, we find 1D invariants with values in braid groups, and 2D invariants with values in $\mathbb{Z}_N$ instead of the expected collection of $N-1$ independent Chern numbers. We provided a detailed pedagogical explanation of how this arises from the mathematical phenomenon of the action of $\pi_1$ on $\pi_2$. We interpreted these classification results in terms of Weyl points and exceptional rings, and connected them to a previously reported nodal braiding in non-Hermitian systems. We illustrate these results using the familiar physics of nematic liquids, and also describe connections to fragile topology of real Hermitian models. Finally, we describe how these invariants are computed, and we illustrate this in a simple computational model.

As we have explained, models representing any of the reported classes can easily be constructed using Weyl points and exceptional rings. This would allow the creation of lattice models in order to experimentally probe edge state phenomena associated with these novel invariants, e.g. in optical lattices or synthetic dimension lattices \cite{Yuan:2018}. Then one could extend the bulk-edge correspondence to this generalized classification and novel invariants, providing a clearer understanding of the bulk-edge correspondence in non-Hermitian systems.

\appendix 

\section{Homotopy groups of $(S^2 \times S^1) / \mathbb{Z}_2$}\label{sec:fiberbundles}
In this appendix, we study the maps $\Pi_1$ and $\Pi_2$ introduced in Sec.~\ref{sec:nonh-class} and use them to formalize some claims about the homotopy groups of $X = (S^2 \times S^1) / \mathbb{Z}_2$. In particular, we can gain some intuition about the homotopy groups by relating $X$ to $S^1$ and to $\mathbb{R}P^2$. Furthermore, the covering space structure alone is insufficient to completely determine $\pi_1(X)$, whereas these calculations do determine $\pi_1(X)$. 

Recall that $\Pi_1 : X \to S^1$ and $\Pi_2 : X \to \mathbb{R}P^2$ are the natural projections from $X$. Both $S^1$ and $\mathbb{R}P^2$ have necessarily forgotten the ordering of the eigenvalues and eigenvectors. Note that once one fixes an ordering on the eigenvectors, an ordering is also determined on the eigenvalues (and vice versa). Thus the map $\Pi_2$ is a fiber bundle with fiber $S^1$, and the map $\Pi_1$ is a fiber bundle with fiber $S^2$. We can formally write this as
\begin{equation}
    \begin{tikzcd}
    S^2 \arrow[r] & X \arrow[r, "\Pi_1"] & S^1 \\
    S^1 \arrow[r] & X \arrow[r, "\Pi_2"] & \mathbb{R}P^2
    \end{tikzcd}
\end{equation}
From a general property of fiber bundles, we obtain the long exact sequences~\cite{Hatcher:2002}
\begin{equation}
    \begin{tikzcd}
    \ldots \arrow[r] & \pi_m(S^2) \arrow[r] & \pi_m(X) \arrow[r] & \pi_m(S^1) \\
    \arrow[r] & \pi_{m-1}(S^2) \arrow[r] & \ldots
    \end{tikzcd}\label{eqn:exact-seq-1}
\end{equation}
\begin{equation}
    \begin{tikzcd}
    \ldots \arrow[r] & \pi_m(S^1) \arrow[r] & \pi_m(X) \arrow[r] & \pi_m(\mathbb{R}P^2) \\
    \arrow[r] & \pi_{m-1}(S^1) \arrow[r] & \ldots
    \end{tikzcd}\label{eqn:exact-seq-2}
\end{equation}
It follows from the exactness of the sequence in Eq.~(\ref{eqn:exact-seq-1}) and from $\pi_1(S^2) = \triv =\pi_0 (S^2)$ that $\pi_m(X) \to \pi_m(S^1)$ is an isomorphism for $m = 1$, therefore the one-dimensional invariant is given by the winding of the eigenvalues in $\UConf_2(\mathbb{C})$. Furthermore, we find using the exact sequence in Eq.~(\ref{eqn:exact-seq-2}) and using $\pi_{2,3,\ldots}(S^1) = \triv$ that $\pi_m(X) \to \pi_m(\mathbb{R}P^2)$ is an isomorphism for $m > 2$. 
For $m = 2$, this map is still an isomorphism because the map $\pi_1(S^1) = \mathbb{Z} \to \pi_1(X) = \mathbb{Z}$ is the inclusion of the even integers. This observation also implies that for $m = 1$, the map is a reduction modulo 2, i.e.~$\pi_1(\mathbb{R}P^2) = \mathbb{Z}_2$ remembers the parity of the winding of the unordered eigenvalues.

We remark that a covering space is a special case of a fiber bundle, one whose fiber is discrete. Applying the long exact sequence to $S^2 \times S^1 \to (S^2 \times S^1) / \mathbb{Z}_2$ reproduces the result that the covering map induces an isomorphism in $\pi_2$ and an inclusion in $\pi_1$, and in fact tells us that $\pi_1(X)$ surjects onto $\mathbb{Z}_2$ with kernel $\pi_1(S^2 \times S^1) = \mathbb{Z}$. However, one is unable to determine the extension type from this information alone, which is one reason it is beneficial to study the fiber bundles $\Pi_1$ and $\Pi_2$.

\begin{acknowledgments}
We thank A.~Tiwari for helpful discussions, and R.-J. Slager and A. Tiwari for providing a feedback on early version of our manuscript. While we were preparing this manuscript, a similar work was brought to our attention (Ref.~\cite{li:2019}) which arrives at the same classification result; our paper in addition provides a pedagogical derivation and many intuitive ways to understand the classification result, in addition to an algorithm for computing the new invariants. 

C.~C.~W. and S.~F. were supported by a Vannevar Bush Faculty Fellowship (Grant No. N00014-17-1-3030) from the U.S. Department of Defense, and by the National Science Foundation Grant No. CBET-1641069. X.-Q.S. was supported DOE Office of Science, Office of High Energy Physics under Grant NO. DE-SC0019380. T.~B. was supported by the Gordon and Berry Moore Foundation's EPiQS Initiative, Grant GBMF4302, and by the Ambizione Program of the Swiss National Science Foundation.

\end{acknowledgments}

\bibliography{bib}{}
\bibliographystyle{apsrev4-1}  
\end{document}


\title{Supplemental Material for:\texorpdfstring{\\}{} Topological Invariants of Non-Hermitian Hamiltonians}

\author{Charles C. Wojcik$^1$}\email[Corresponding author: ]{cwojcik@stanford.edu}
\author{Xiao-Qi Sun$^{2,3}$}
\author{Tom\'{a}\v{s} Bzdu\v{s}ek$^{2,4,5}$}
\author{Shanhui Fan$^1$}

\affiliation{$^{1}$Department of Electrical Engineering, Ginzton Laboratory, Stanford University, Stanford, CA 94305, USA}
\affiliation{$^{2}$Department of Physics, McCullough Building, Stanford University, Stanford, CA 94305, USA}
\affiliation{$^{3}$Stanford Center for Topological Quantum Physics, Stanford University, Stanford, CA 94305, USA}
\affiliation{$^{4}$Condensed Matter Theory Group, Paul Scherrer Institute, CH-5232 Villigen PSI, Switzerland}
\affiliation{$^{5}$Department of Physics, University of Z\"{u}rich, 8057 Z\"{u}rich, Switzerland}

\date{\today}

\maketitle

\section{A road-map to the \texorpdfstring{$\bs{k}\cdot\bs{p}$}{k.p} model}

In this section, we present a $\bs{k}\cdot\bs{p}$ Hamiltonian that exhibits a  non-trivial braiding of Weyl point as one free parameter is tuned. The discussion is split in two parts. First, in Sec.~\ref{eqn:kp-simple} we obtain a Hamiltonian that exhibits a single exceptional nodal line. This is achieved by starting with a simple Weyl Hamiltonian and by including a $\bs{k}$-independent non-Hermitian perturbation. In Sec.~\ref{eqn:kp-complicated} we include in the developed model an additional and more complicated $\bs{k}$-dependent non-Hermitian perturbation. After properly adjusting the parameters, we find that the resulting model exhibits two Weyl points that are non-trivial braided around the exceptional nodal line.

\subsection{Hamiltonian with a single exceptional nodal line}\label{eqn:kp-simple}

We begin with the simplest Weyl point Hamiltonian
\begin{equation}
\mcH^{(0)}(\bs{k}) = \bs{k}\cdot\bs{\sigma}, \label{eqn:Weyl-simple}
\end{equation}
where $\bs{\sigma}=(\sigma_x,\sigma_y,\sigma_z)$ is the vector of Pauli matrices. The spectrum of the Weyl Hamiltonian is $\varepsilon^0_\pm(\bs{k}) = \pm\abs{\bs{k}}$. The superscript of the Hamiltonian indicates the position of the Weyl point inside the complex $k_x + \imi k_y$ plane. This convention will become useful later in Sec.~\ref{eqn:kp-complicated}.

Non-Hermitian perturbations generically ``inflate'' a Weyl point into a ring of exceptional nodes. Here, we consider the simple $\bs{k}$-independent non-Hermitian perturbation $\delta \mcH = \imi \tfrac{m}{2}\sigma_y$ with $m\in\reals$. The combined Hamiltonian 
\begin{equation}
\mcH^{(0)}(\bs{k})+\delta\mcH \equiv \mcH(\bs{k}) = k_x\sigma_x + (k_y + \imi\tfrac{m}{2})\sigma_y + k_z \sigma_z \label{eqn:pert-Weyl-Hamilt}
\end{equation} 
has energy spectrum
\begin{equation}
\varepsilon_\pm(\bs{k}) = \sqrt{\bs{k}^2 + \imi m k_y - \tfrac{m^2}{4}},
\end{equation}
which exhibits an exceptional degeneracy at $k_y = 0$ along a ring $4(k_x^2 + k_z^2) = m^2$. 

We shift the momentum coordinates as $\bs{k} = \bs{g} + \delta\bs{k}$, where $\bs{g} = (\tfrac{m}{2},0,0)$. Furthermore, we find it useful to instead work with matrices $\sigma_\pm = \sigma_x \pm \imi \sigma_y$ (together with $\sigma_z$) and with momentum-component combinations $k_\pm = \delta k_x \pm \imi \delta k_y$ (together with $\delta k_z = k_z$). The inverse transformations are
\begin{equation}
\delta k_x = \frac{k_+ + k_-}{2}\quad \textrm{and} \quad \delta k_y = \frac{k_+ - k_-}{2\imi},
\end{equation}
and similarly for the Pauli matrices $\sigma_{x}$ and $\sigma_y$. We will use the fact that the Hamiltonian of the form
\begin{equation}
\mcH = h_+\sigma_+ + h_-\sigma_- + h_z\sigma_z
\end{equation}
has energy eigenvalues 
\begin{equation}
\varepsilon(\bs{k})\pm = \pm \sqrt{4 h_+ h_- + h_z^2}. \label{eqn:sigma+-spect}
\end{equation}
We also remark that the Weyl Hamiltonian in Eq.~(\ref{eqn:Weyl-simple}) can be rewritten as 
\begin{equation}
\mcH^{(0)}(\bs{k}) = \tfrac{1}{2}\left(k_+ \sigma_- + k_- \sigma_+\right)  + k_z \sigma_z. \label{eqn:Weyl-simple+-basis}
\end{equation}
This decomposition will come in handy later in Sec.~\ref{eqn:kp-complicated}.

In the basis of $\sigma_\pm$ and using the momentum coordinates $k_\pm$, the perturbed Weyl Hamiltonian from Eq.~(\ref{eqn:pert-Weyl-Hamilt}) becomes 
\begin{eqnarray}
\mcH(\bs{k}) &=& \left(\tfrac{m}{2} + \delta k_x\right)\sigma_x + \left(\delta k_y +\imi \tfrac{m}{2}\right)\sigma_y + \delta k_z \sigma_z \\
&=&\tfrac{1}{2}\left[(m + k_-) \sigma_+ + k_+ \sigma_-\right]+ k_z \sigma_z.
\end{eqnarray}
Near the origin of coordinates, we can approximate $m + k_- \approx m$, and we further set $m=1$. This approximates the Hamiltonian to
\begin{equation}
\mcH(\bs{k}) = \tfrac{1}{2}\left(\sigma_+ + k_+ \sigma_-\right) + k_z \sigma_z \label{eqn:k.p-simple}
\end{equation}
with spectrum
\begin{equation}
\epsilon_\pm = \pm \sqrt{k_+ + k_z^2}\label{eqn:ener-scaling},
\end{equation}
which exhibits an exceptional nodal line inside the $k_y = 0$ plane along $k_x = - k_z^2$. The Hamiltonian in Eq.~(\ref{eqn:k.p-simple}) provides the starting point for the more complicated Hamiltonian discussed in the next subsection. One could make the exceptional nodal line straight, pointing along $k_x = 0 = k_y$, by dropping the $k_z \sigma_z$ term. However, we keep this term in the Hamiltonian, because it is very important for the generation of Weyl points inside the $k_z = 0$ plane in the next section.

\subsection{Hamiltonian with additional Weyl points}\label{eqn:kp-complicated}

Our next goal is to supplement the Hamiltonian in Eq.~(\ref{eqn:k.p-simple}) with a perturbation that produces Weyl points at well-controlled positions. More specifically, we want a pair of Weyl points moving inside the $k_z = 0$ plane along semicircles $k_+ = r \e{\pm \imi \phi}$ with some radius $r$ and polar angle $\phi\in[0,\pi]$. The semicircles ``circumnavigate'' the exceptional line inside the $k_z = 0$ plane. Our strategy to get such a model is to consider $k_z$-independent perturbation, and temporarily limit our attention to the spectrum inside the $k_z = 0$ plane. The only $k_z$ dependence of our model would come from the $k_z \sigma_z$ term that is readily present in Eq.~(\ref{eqn:k.p-simple}), or even in Eq.~(\ref{eqn:Weyl-simple}).

Inside the $k_z = 0$ plane, both the Weyl points as well as the cross-section of the exceptional nodal line appear as point-like objects. However, they are associated with a different topological structure. While the cross-section of the exceptional nodal line at $k_x = 0 = k_y$ is associated with a non-trivial winding of the complex-valued band energies [corresponding to the first homotopy group $\pi_1(M)$], the Weyl points do not exhibit such winding. Since we want the perturbed Hamiltonian to exhibit just the single exceptional nodal line at the origin of coordinates, we should keep the 
\begin{equation}
\eps_\pm = \sqrt{4h_+ h_-} \propto \pm \sqrt{k_+}\label{eqn:energies-scaling}
\end{equation}
scaling of the band energies, as present in Eq.~(\ref{eqn:ener-scaling}).

It follows from shifting the coordinates in Eq.~(\ref{eqn:Weyl-simple}) that a Hamiltonian with a single Weyl point at position $k_+ = r \e{\imi \phi}$ (which by complex conjugation corresponds to $k_- = r\e{-\imi \phi}$) is simply 
\begin{equation}
\mcH^{(r\e{\imi\phi})}\!(\bs{k}) \!=\! \tfrac{1}{2}\!\left[(k_+ \!\!-\! r\e{\imi \phi})\sigma_- \!+\! (k_- \!\!\!-\! r\e{-\imi\phi})\sigma_+\!\right]\!\!+\!k_z \sigma_z.\! \label{eqn:Weyl-shifted}
\end{equation}
By referring to Eq.~(\ref{eqn:sigma+-spect}), it is clear that the spectrum of the model in Eq.~(\ref{eqn:Weyl-shifted}) is $\eps_\pm = \pm\sqrt{\abs{k_+ - r\e{\imi\phi}}^2 + k_z^2}$, as expected for a Weyl point at that position. We also emphasize that Eq.~(\ref{eqn:Weyl-shifted}) explicitly adopts the convention outlined below Eq.~(\ref{eqn:Weyl-simple}), namely that the superscript of the Weyl Hamiltonian indicates the position of the Weyl point inside the $k_z = 0$ plane.

To get a pair (and potentially even more) of Weyl points inside the $k_z = 0$ plane, we just need to keep adding factors $(k_\pm - r_i\e{\pm \varphi_i})$ to $\sigma_\pm$. However, there is actually more than one way to do this. For example, there are two ways to get a Hamiltonian with two in-plane Weyl points located at $r_1\e{\imi\varphi_1}\equiv \hat{z}_1$ and at $r_2\e{\imi\varphi_2} = \hat{z}_2$, namely
\begin{eqnarray}
\mcH^{(\hat{z}_1,\hat{z}_2)}_a\!(\bs{k})\!&=&\! \tfrac{1}{2}\!\left[(k_+ \!-\! \hat{z}_1)(k_+ \!-\! \hat{z}_2)\sigma_- \!+\! \textrm{h.c.}\right]\!+\!k_z \sigma_z \\
\mcH^{(\hat{z}_1,\hat{z}_2)}_b\!(\bs{k}) \!&=&\! \tfrac{1}{2}\!\left[(k_+ \!-\! \hat{z}_1)(k_- \!-\! {\hat{z}_2}^*)\sigma_- \!+\! \textrm{h.c.}\right]\!+\!k_z \sigma_z, \label{eqn:two-Weyls-b}
\end{eqnarray}
where ``h.c.'' stands for ``Hermitian conjugate", and the asterisk ``$^*$'' indicates complex conjugation. 

The Hamiltonians in Eqs.~(\ref{eqn:Weyl-shifted}--\ref{eqn:two-Weyls-b}) are all Hermitian, and do not exhibit a non-trivial winding of the complex-valued band energies. Therefore, they do not exhibit the scaling of band energies expressed by Eq.~(\ref{eqn:energies-scaling}) in the limit of large $\abs{k_\pm}$. Note also that Hamiltonians with two (or more generally with $n$) in-plane Weyl points would scale as $\abs{k_\pm}^2$ (resp.~as $\abs{k_\pm}^n$) for large $\abs{k_\pm}$, i.e.~this term will at large momenta take over the exceptional-line Hamiltonian given by Eq.~(\ref{eqn:k.p-simple}), which we intend to perturb. To make sure that the perturbed Hamiltonian would retain the correct band energy winding for $\abs{k_\pm}\to\infty$, we multiply the term proportional to $\sigma_-$ by an additional factor of $k_+$. 

We now collect all that we learned from the previous discussion. We opt for the variant $b$ of including two in-plane Weyl points, corresponding to Eq.~(\ref{eqn:two-Weyls-b}). Since we want the two Weyl points to be located at polar angles $\pm\alpha$, we perturb the exceptional-line Hamiltonian in Eq.~(\ref{eqn:k.p-simple}) with 
\begin{eqnarray}
\mcH'(\bs{k};\alpha) &=& + \; (k_- + \e{-\imi\alpha})(k_+ + \e{-\imi\alpha})\sigma_+ \nonumber \\
&\phantom{=}& + \; (k_+ + \e{+\imi\alpha})(k_- + \e{+\imi\alpha})k_+\sigma_-.\label{eqn:final-pert}
\end{eqnarray}
The combination $\mcH(\bs{k}) + \mcH'(\bs{k};\alpha)$ of the Hamiltonians in Eqs.~(\ref{eqn:k.p-simple}) and~(\ref{eqn:final-pert}) corresponds to the model presented in Eq.~(1) of the main text. This model exhibits two in-plane Weyl points at complex-conjugated values of $k_+$, and a single exceptional nodal line crossing the $k_z = 0$ plane at $k_x = 0 = k_y$. 

It can be analytically derived that the Weyl points of the constructed Hamiltonian move on a circle with radius $r = 1/\sqrt{2}$, and that their polar coordinates obey $\cos\phi = -\sqrt{2}\cos\alpha$. It is therefore clear that the Weyl points exist inside the $k_z = 0$ plane only for $\alpha\in (\tfrac{\pi}{4},\tfrac{3\pi}{4})$. A more careful analysis reveals that the Weyl points are \emph{ejected} from the exceptional line touching the $k_z = 0$ plane as one increases $\alpha$ through $\pi/4$, and that the two Weyl points pairwise annihilate as one increases $\alpha$ through $3\pi/4$. 

\section{The alternative lattice model}

As an alternative model that could be more easily implemented in experiments, we consider the lattice model
\begin{eqnarray}
\mcH(\bs{k};m)  
&=& \e{\imi \frac{k_z}{2}}\left[\cos\left(\tfrac{k_z}{2}-\tfrac{\pi}{4}\right)\sin k_x \sigma_x + \right. \nonumber \\ 
&\phantom{=}&+\left.\cos\left(\tfrac{k_z}{2}+\tfrac{\pi}{4}\right)\sin k_y \sigma_y\right. \label{eqn:alt-lattice-model} \\
&\phantom{=}&+ \left. \left(\sin k_z \cos\tfrac{k_z}{2} - 2m \sin \tfrac{k_z}{2} \right)\sigma_z \right]. \nonumber
\end{eqnarray}
This model respects the periodicity of the momentum space, and the complex-valued determinant $\det\mcH(\bs{k};m)$ exhibits a non-trivial winding along the $k_z$ direction of the Brillouin zone torus. More explicitly, the $\e{\imi \tfrac{k_z}{2}}$ prefactor in Eq.~(\ref{eqn:alt-lattice-model}) guarantees that $\textrm{arg}[\det\mcH(\bs{k};m)] = k_z$. 

The non-Hermitian lattice model in Eq.~(\ref{eqn:alt-lattice-model}) exhibits several Weyl points. First, for all values of $m$ there are Weyl points located at $\bs{k}=(0,0,0)$, $(0,\pi,0)$, $(\pi,0,0)$, $(\pi,\pi,0)$ inside the $k_z = 0$ plane of the Brillouin zone. Furthermore, for $m\in(0,1)$, there are additional eight Weyl points located at the same $k_x$ and $k_y$ as the previous quadruplet, and with $\sin \tfrac{k_z}{2} = \pm \sqrt{1-m}$. We remark that this model does not exhibit any exceptional lines. Nevertheless, we find that the chirality of the Weyl points [corresponding to $\pi_2(M)$] interacts non-trivially with the winding number along the $k_z$-direction of the Brillouin zone torus [corresponding to $\pi_1(M)$].

Here, we shall focus on the Weyl points on the $k_x=k_y=0$ line, while the analyses of Weyl points on the other lines such as $(k_x,k_y)=(0,\pi),(\pi,0)$ and $(\pi,\pi)$ are similar. First, we would like expand the Bloch Hamiltonian around the Weyl point $k_x=k_y=k_z=0$:
\begin{equation}
    \mcH(\bs{k};m)\approx \frac{1}{\sqrt{2}}k_x\sigma_x+\frac{1}{\sqrt{2}}k_y \sigma_y+(1-m)k_z \sigma_z
\end{equation}
Therefore, for $m<1$, the Weyl point chirality is positive, while for $m>1$, the Weyl point chirality is negative. 

For $m<0$ and $m>1$, there is only one Weyl point on the line at $k_z=0$. However, this Weyl point is of opposite chirality for the situations with $m<0$ resp.~with $m>1$. Therefore, as we as we tune $m$ from a negative number to positive number larger than $1$, the ``total chirality" of Weyl points on the $(k_x,k_y)=(0,0)$ line changes sign without other band nodes moving onto the line. As we have explained in the main text, this phenomenon corresponds to the fact that the total chirality of Weyl points cannot be well-defined globally in the Brillouin zone. 

Here, we would also like to understand the process in more details. As $m$ increases from negative to positive, a pair of Weyl points are created at $k_z=\pi$, which locally exhibit opposite chirality. As $m$ grows from 0 to 1, the two Weyl points move towards $k_z=0$ from the positive $k_z$ side and from the negative $k_z$ side, respectively. Along the fixed $k_x = 0 = k_y$ line, the phase of the two eigenvalues of the Bloch Hamiltonian are $\pm e^{i k_z/2}$, which means that the two energy bands exchange as one increases $k_z$ by $2\pi$, similar to encircling an exceptional line. Therefore, as the two Weyl points meet at $k_z=0$, they effectively ``encircle an exceptional line" and now have the same chirality, as illustrated in Fig.~1 of the main text. The extra chirality carried by the two Weyl points from $k_z=\pi$ hence flips the chirality of the Weyl point at $k_z=0$. We illustrate the exchange of Weyl points of the model in Eq.~(\ref{eqn:alt-lattice-model}) in Fig.~\ref{fig:my_label} of the Supplemental Material.
\begin{figure}
    \centering
    \includegraphics[width=0.5\textwidth]{Fig1S.pdf}
    \caption{Weyl points of the model in Eq.~(\ref{eqn:alt-lattice-model}), which are located inside (a part of) the $k_y=0$ plane for (a) $m<0$, (b) $m=0.25$ and (c) $m>1$. We find that for $m<0$, there is a Weyl point (red) of positive chirality at $k_x=k_z=0$. A pair of Weyl points (green) are created at $(k_x,k_z)=(\pi,0)$ for $m=0$. After bringing the two green Weyl points to $k_z = 0$, they merge with the Weyl point located at $k_x=k_z=0$. There remains a Weyl point (blue) with negative chirality at $k_x=k_z=0$ for $m>1$. 
    }
    \label{fig:my_label}
\end{figure}

\section{Topological invariants}

In this section, we complete the details of computing the action $\triangleright$ of $\pi_1(M)$ on $\pi_2(M)$, which are omitted in the main text. Here, $M$ is the space of $2\times 2$ Hamiltonians that are traceless and have spectrum normalized to absolute value $1$. We shall first review the formalism introduced by the main text. First, we express the space $M$ as a coset space $\mathsf{G}/\mathsf{H}$, where $\mathsf{G}$ is a simply connected Lie group and $\mathsf{H}$ is the stabilizer subgroup. Then, we use the coset expression and the computational algorithm described in Ref.~\cite{Mermin:1979} to derive $\pi_1(M)$ and $\pi_2(M)$, as well as the action $\triangleright$.

As explained in the main text, we can identify any Hamiltonian in $M$ using $(V,t)\in \mathsf{SL}(2,\cmplx)\times \reals \equiv \mathsf{G}$, as expressed by Eq.~(2) of the main text. Furthermore, the main text argues that the stabilizer group $\mathsf{H}$ consists of the following elements in $\mathsf{G}$: $(R(z),n)$ for even $n$, and $(i\sigma_y\cdot R(z),n)$ for odd $n$, where $R(z)=\text{diag}(z,1/z)$ with $z$ being any complex number except of zero (which we indicate as $\cmplx/\{0\} \equiv \cmplx^\times$). We then use a mathematical theorem from Ref.~\cite{Mermin:1979} to show that $\pi_{2}(M)=\pi_{1}(\mathsf{H})=\intg$ [i.e.~the Chern number on a 2-sphere corresponds to ``looping'' of the argument of $R(z)$ around the origin of $\cmplx^\times$], and $\pi_{1}(M)=\pi_{0}(\mathsf{H})=\intg$ [i.e.~the winding number on a 1-sphere corresponds to the connected component $n$ of the stabilizer group $\mathsf{H}$]. 

According to Fig.~4(d) in the main text, we should study the conjugation of elements in $\pi_{2}(M)$ by elements in $\pi_{1}(M)$. The elements in $\pi_{2}(M)$ are represented as topologically distinct loops in the space of $\mathsf{H}$ ($\pi_{1}(\mathsf{H})$). The loops can be parameterized by $T_{n_2} (z_2)=(R(z_2),n_2)$ for even $n_2$ and and $T_{n_2} (z_2)=(i\sigma_y\cdot R(z_2),n)$ for odd $n_2$, where $z_2$ is taken along a path that loops around the origin of the complex plane $\cmplx^\times$. 
On the other hand, the elements in $\pi_{1}(M)$ are represented as disconnected points in the space of $\mathsf{H}$ ($\pi_{0}(\mathsf{H})$). This corresponds to $T_{n_1}(z_1)$ with some $z_1\in\cmplx^\times$. Without loss of generality, we set $z_1 = 1$ in our arguments below (we comment on the case of general $z_1$ in the last paragraph below).

We can now compute $T_{n_1}(1)\circ T_{n_2}(z_2)\circ T_{n_1}(1)^{-1}$ explicitly. Recall that the group $\mathsf{G}$ is a direct product of Abelian part of integer number addition $\reals$ and of a non-Abelian part of $\mathsf{SL}(2,\cmplx)$. The Abelian part of $T_{n_2}(z_2)$ does not change upon conjugation by $T_{n_1}(1)$. Therefore, we only need to calculate the non-Abelian part of $T_{n_1}(1)\circ T_{n_2}(z_2)\circ T_{n_1}(1)^{-1}$. The calculation has to be split into several cases, corresponding to different parities of $n_1$ and $n_2$. 

For even $n_1$, $T_{n_1}(1)=(I_{2\times 2},n)$ commute with $T_{n_2}(z_2)$ and therefore
\begin{equation}
T_{n_1}(1)\circ T_{n_2}(z_2)\circ T_{n_1}(1)^{-1}=T_{n_2}(z_2), \label{eqn:action-n1-even}
\end{equation}
On the other hand, for odd $n_1$, we can use the following commutating relations:
\begin{equation}
\begin{split}
    &i\sigma_y\cdot R(z_2)=R(1/z_2)\cdot i\sigma_y\\
    &i\sigma_y\cdot (i\sigma_y\cdot R(z_2))=(i\sigma_y\cdot R(1/z_2))\cdot i\sigma_y
\end{split}
\end{equation}
to derive that for any $n_2$, we have
\begin{equation}
    T_{n_1}(1)\circ T_{n_2}(z_2)=T_{n_2}(1/z_2)\circ T_{n_1}(1).
\end{equation}
It follows that
\begin{equation}
    T_{n_1}(1)\circ T_{n_2}(z_2)\circ T_{n_1}(1)^{-1}=T_{n_2}(1/z_2).\label{eqn:action-n1-odd}
\end{equation}
The results in Eqs.~(\ref{eqn:action-n1-even}) and~(\ref{eqn:action-n1-odd}) can be compactly unified into a single equation
\begin{equation}
    T_{n_1}(1)\circ T_{n_2}(z_2)\circ T_{n_1}(1)^{-1}=T_{n_2}\left(z_2^{P(n_1)}\right),
\end{equation}
where $P(n_1)=\pm 1$ is the parity of $n_1$. Note that $1/z_2$ has opposite ``looping'' around the origin of $\cmplx^\times$ than $z_2$. 
It follows that that for odd $n_1$ [odd elements of $\pi_1(M)$], the conjugation flips the sign of the $\pi_2(M)$ charge.

For a general $z_1$, one can explicitly compute for the four different combinations of parities of $n_1$ and $n_2$ the following results:
\begin{itemize}
\item If $n_{1}$ is even and $n_{2}$ is even, then
\begin{equation}
    T_{n_1}(z_1)\circ T_{n_2}(z_2)\circ T_{n_1}(z_1)^{-1}=T_{n_2}(z_2).
\end{equation}
\item If $n_{1}$ is even and $n_{2}$ is odd, then
\begin{equation}
    T_{n_1}(z_1)\circ T_{n_2}(z_2)\circ T_{n_1}(z_1)^{-1}=T_{n_2}(z_2/z_1^2).
\end{equation}
\item If $n_{1}$ is odd and $n_{2}$ is even, then
\begin{equation}
    T_{n_1}(z_1)\circ T_{n_2}(z_2)\circ T_{n_1}(z_1)^{-1}=T_{n_2}(1/z_2).
\end{equation}
\item If $n_{1}$ is odd and $n_{2}$ is odd, then
\begin{equation}
    T_{n_1}(z_1)\circ T_{n_2}(z_2)\circ T_{n_1}(z_1)^{-1}=T_{n_2}(z_1^2/z_2).
\end{equation}
\end{itemize}
The three equations are compactly summarized in Eq.~(4) of the main text.

\bibliography{bib}{}
\bibliographystyle{apsrev4-1}